# Topological connection between vesicles and nanotubes in single-molecule lipid membranes driven by head-tail interactions

Niki Baccile,[a,*] Cédric Lorthioir,[a] Abdoul Aziz Ba,[a] Patrick Le Griel,[a] Javier Perez,[b] Daniel Hermida-Merino,[c,d] Wim Soetaert,[e] Sophie L. K. W. Roelants[e]

[a] Sorbonne Université, Centre National de la Recherche Scientifique, Laboratoire de Chimie de la Matière Condensée de Paris, LCMCP, Paris, 75005, France

[b] Synchrotron Soleil, L'Orme des Merisiers, Saint-Aubin, BP48, Gif-sur-Yvette Cedex, 91192, France

[c] Netherlands Organisation for Scientific Research (NWO), DUBBLE@ESRF BP CS40220, Grenoble, 38043, France

[d] Departamento de Física Aplicada, CINBIO, Universidade de Vigo, Campus Lagoas-Marcosende, Vigo, 36310, Spain

[e] InBio, Department of Biotechnology, Ghent University, Ghent, 9000, Belgium

**\* Corresponding author:**
Dr. Niki Baccile
E-mail address: niki.baccile@sorbonne-universite.fr
Phone: +33 1 44 27 56 77

**Abstract**

Lipid nanotube-vesicle networks are important channels for intercellular communication and transport of matter. Experimentally observed in neighboring mammalian cells, but also reproduced in model membrane systems, a broad consensus exists on their formation and stability. Lipid membranes must be composed of at least two molecular components, each stabilizing low (generally a phospholipid) and high curvatures. Strong anisotropy or enhanced conical shape of the second amphiphile is crucial for the formation of nanotunnels. Anisotropic driving forces generally favor nanotube protrusions from vesicles. In the present work, we report the unique case of topologically-connected nanotubes-vesicles obtained in the absence of directional forces, in single-molecule membranes, composed of an anisotropic bolaform glucolipid, above its melting temperature, $T_m$. Cryo-TEM and fluorescence confocal microscopy show the interconnection between vesicles and nanotubes in a single-phase region, between 60° and 90°C under diluted conditions. Solid-state NMR demonstrates that the glucolipid can assume two distinct configurations, head-head and head-tail. These arrangements, seemingly of comparable energy above the $T_m$, could explain the existence and stability of the topologically-connected vesicles and nanotubes, which are generally not observed for classical single-molecule phospholipid-based membranes above their $T_m$.

**Introduction**

Topological connections between closed lipidic compartments through nanotubes[1–3] have been shown to play a crucial role in the transfer of matter and communication in neighboring mammalian cells.[4] These singular nanosystems, observed since the 1990s as spontaneous non-equilibrium structures in electroformed model liposome membranes,[5,6] have since been largely studied, both experimentally and theoretically.[7–9] Addressed in the literature by different terms, tunnelling nanotubes (TNT),[1,4,9] block liposomes[10–12] or nanotube-vesicle networks,[13–16] (instead of tubes, some work speaks of tethers[6,17]) all refer to a similar phenomenon, driven by various internal or external forces. The latter must overcome the energy barrier needed to bend a phospholipid bilayer from low positive mean and Gaussian curvatures (vesicle) to a high mean and zero Gaussian (tube) curvatures.

A large body of both experimental and theoretical work has shown that budding and eventual nanotube formation from an existing membrane can only occur spontaneously for membranes of at least two molecular components[9–12,18–26] and below phase transition event (corresponding to $T_m$ for lipids or glass transition temperature, $T_g$, for block copolymers).[27–29] If internal driving forces, like intra-membrane polymer-polymer phase separation,[30] can trigger nanotube protrusion, external anisotropic forces like electroformation,[5–8,31,32] osmotic pressure,[6] laser "tweezers"[33] or electrodynamics[13–16] must generally be employed with, in some cases, an impressive degree of 2D and 3D organization.[13–16] The origin of spontaneous nanotube formation has been shown to be related to a nanoscale phase separation between two membrane molecules, stabilizing low and high curvatures respectively,[10–12,34] with at least one molecule being highly anisotropic.[24,35,36] Theoretically, this behavior has been explained by deviations in the elastic properties of membranes due to in-plane orientational ordering of membrane inclusions composed of anisotropic amphiphiles, these referring to a non-symmetrical shape upon a 90° tilt along the amphiphile axis.[8,9,19,20,23,36]

In this work, we show unexpected nanotubing of membranes prepared from a single-molecule lipid, in the absence of external directional forces and above the lipid's $T_m$. This phenomenon is observed for a novel anisotropic double amphiphile (bolaform amphiphile, or bolaamphiphile), a glucolipid composed of β-D-glucose and a C18:1-*cis* fatty alcohol (G-C18:1-OH, Figure 1). This compound is obtained by microbial fermentation of a genetically-modified *S. bombicola* yeast in the presence of oleyl alcohol[37] and is developed in the broader context of extending the library of new biobased surfactants and lipids, in view of replacing petrochemical low molecular weight amphiphiles.[38–44] The structure of G-C18:1-OH is analogous to that of other microbial glycolipids developed through genetic engineering.[45,46]

Topological connections between nanotubes and vesicles are observed by means of cryogenic transmission electron microscopy (cryo-TEM), fluorescence microscopy and temperature-resolved *in situ* small and wide angle X-ray scattering (SAXS-WAXS) above the melting temperatures, $T_m$= 48.3°C. Spin diffusion 2D solid-state nuclear magnetic resonance (ssNMR) spectroscopy under magic angle spinning (MAS) help understanding the vesicle-nanotube coexistence. The bolaform glucolipids could be in a head-head/tail-tail configuration in the vesicles and in a head-tail configuration in the nanotubes. These facts could explain the stability of nanotubes, while the following hypothesis is formulated for their formation: membrane inclusions with different orientational ordering[13–16] possibly driven by inter-vesicle collisions.

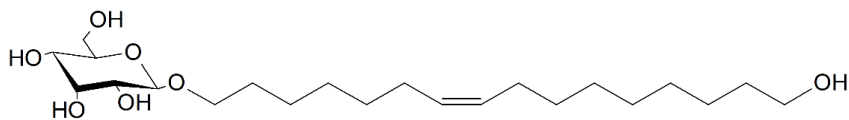

**Figure 1 – Non-acetylated C18:1 alcohol glucoside, G-C18:1-OH, is obtained by a bioprocess performed with modified *S. bombicola* yeast.**

**Experimental Section**

*Synthesis of non-acetylated C18:1 alcohol glucosides (G-C18:1-OH).* G-C18:1-OH (Mw= 418.56 g.mol$^{-1}$) was produced by aerobic whole cell bioprocess with a modified *S. bombicola* strain as described by Van Renterghem et al. (Fig. S4 in Ref. [37]). The molecule was purchased from the Bio Base Europe Pilot Plant (Gent, Belgium) and has the generalized chemical structure given in figure Figure 1. The HPLC and $^1$H NMR spectrum (MeOD-d4) with peak assignment are shown in Figure S 1. High purity levels (99%) and high degree of uniformity were obtained, as can be derived from HPLC-ELSD chromatogram, $^1$H NMR and table of contaminant given in Figure S 1.

*$^1$H solution Nuclear Magnetic Resonance (NMR).* $^1$H solution NMR experiments were performed on a Bruker Avance III 300 spectrometer using a 5 mm $^1$H-X BBFO probe using methanol-d4 as solvent. The number of transients is 8 with 3 s recycling delay, an acquisition time of 5.46 s and a receiver gain of 362. The $^1$H NMR spectrum and relative assignment are shown in Figure S 1. $^{13}$C solution NMR were performed on the same probe using DMSO-d6. Referencing is done with respect to TMS. $\delta_{1H}$= 0 ppm, $\delta_{13C}$= 0 ppm.

*Sample preparation.* The sample was dissolved in milliQ-grade water at the concentration of 5 mg/mL (0.5 wt%), although specific experiments involving SAXS, solid-state NMR or DSC have been performed on hydrated but more concentrated samples (50 wt%). Lack of pH-sensitive probes (e.g., COOH groups) in G-C18:1-OH, as otherwise found in other microbial amphiphiles,[47] but also the will to avoid ion-specific effects,[48] exclude the use of buffer. In all experiments, the sample is heat above 110°C and analyzed during cooling. On the bench qualitative experiments are performed using a CH3-150 Combitherm-2 dry block heating device.

*Differential Scanning Calorimetry (DSC).* DSC was performed using a DSC Q20 apparatus from TA Instruments equipped with the Advantage for Q Series Version acquisition software (v5.4.0). Acquisition was performed on both hydrated and dry powder sample (~ 3-5 mg) sealed in a classical aluminium cup and using an immediate sequence of heating and cooling ramps at a rate of 10°C.min$^{-1}$. Melting temperatures, $T_m$, 1 and 2, $T_{m1}$ and $T_{m2}$, were taken at the minimum of the endothermic peak.

*Small and Wide Angle Scattering experiments.* Small angle neutron scattering (SANS) experiments were performed at the D11 beamline of Institut Laue Langevin (Grenoble, France) during the run No. 9-13-778. Four q-ranges have been explored and merged using the following wavelengths, λ, and sample-to-detector (StD) distances. 1) ultra-low q: λ= 13.5Å, StD= 39 m; 2) low-q: λ= 5.3Å, StD= 39 m; 3) mid-q: λ= 5.3Å, StD= 8 m; 4) high-q: λ= 5.3Å, StD= 1.4 m. The sample (C= 5 mg/mL) was prepared in 99.9% $D_2O$ to limit the incoherent background scattering. The sample solution was analyzed in standard 1 mm quartz cells. The direct beam, an empty quartz cell, and $H_2O$ (incoherent scatterer) within the quartz cell were recorded and boron carbide ($B_4C$) was used as neutron absorber. The sample acquisition was measured at 90°C, where temperature was controlled through the controller thermalized sample holder available at the beamline. The background sample ($D_2O$) signal was subtracted from the experimental data. Absolute values of the scattering intensity were obtained from the direct determination of the number of neutrons in the incident beam and the detector cell solid angle. The 2D raw data were corrected for the ambient background and empty cell scattering and normalized to yield an absolute scale (cross section per unit volume) by the neutron flux on the samples. The data were then circularly averaged to yield the 1D intensity distribution, I(q). The software package Grasp (developed at ILL and available free of charge) was used to integrate the data, while the software package SAXSUtilities (developed at ESRF and available free of charge) was used to merge the data acquired at all configurations and subtract the background.

Wide-angle X-ray scattering (WAXS) was performed under temperature control at the SWING beamline of SOLEIL synchrotron facility (Saint-Aubin, France) during the run number 20201747 (energy: 14 keV, sample-to-detector distance: 0.5 m). Sample concentration in $H_2O$ was C= 5 mg/mL. The 2D data were integrated azimuthally at the beamline using the software Foxtrot and in order to obtain the *I(q) vs. q* spectrum after masking the beam stop shadow. Silver behenate ($d_{(001)}$= 58.38 Å) was used as a standard to calibrate the q-scale. Sample solutions were inserted in borosilicate capillaries of 1.5 mm in diameter. Capillaries were flame-sealed. A capillary oven with controlled temperature (± 0.5°C), provided at the beamline, was used to control the sample temperature. Experiments are recorded between 111°C and 25°C. Data were normalized by the transmission and calibrated to the SAXS signal of $H_2O$ at large q-values (I= 0.0163 cm$^{-1}$) in order to obtain an absolute intensity scale. The water signal was measured by subtracting the signal of the empty capillary from the signal of a water-filled capillary. The signal of (water + capillary) was used as background for the samples and it was subtracted after integration of the 2D data.

Temperature-resolved *in situ* small and wide-angle X-ray scattering (SAXS-WAXS) was performed at the DUBBLE beamline of ESRF synchrotron facility (Saint-Aubin, France) during the run number SC-5125 (energy: 12 keV, sample-to-detector distance: 2.6 m). Sample concentration in $H_2O$ was C= 50 wt%. Silver behenate ($d_{(001)}$ = 58.38 Å) and α-$Al_2O_3$ ($d_{(012)}$ = 3.48 Å) are respectively used as SAXS and WAXS standards to calibrate the q-scale. The sample was inserted in a flame-sealed borosilicate capillary of 2 mm diameter placed inside a Linkam cell with controlled temperature (± 0.5°C) and coupled to the beamline. Data were normalized by the transmission but not in an absolute intensity scale. The signal of (water + capillary) was used as background for the samples and it was subtracted after integration of the 2D data. SAXS-WAXS data are recorded both during heating from room temperature (RT) to 110°C and cooling from 110°C to RT at a rate of 1°C/min. The rate was determined empirically and it was found to be a good compromise to obtain a stable SAXS-WAXS signal between heating (5-10 min per temperature) and cooling (< 1 min per temperature) experiments.

*Cryogenic transmission electron microscopy (cryo-TEM).* Cryo-TEM experiments were carried out on an FEI Tecnai 120 twin microscope operating at 120 kV and equipped with a Gatan Orius CCD numeric camera. The sample holder was a Gatan Cryoholder (Gatan 626DH, Gatan). Digital Micrograph software was used for image acquisition. Cryofixation was performed using a homemade cryofixation device. The solutions were deposited on a glow-discharged holey carbon coated TEM copper grid (Quantifoil R2/2, Germany). Excess solution was removed and the grid was immediately plunged into liquid ethane at -180°C before transferring them into liquid nitrogen. All grids were kept at liquid nitrogen temperature throughout all experimentation. Cryo-TEM images have been treated and analyzed using Fiji (is just ImageJ) software, available free of charge at the developer's website.[49]

The sample's concentration for cryo-TEM was C= 5 mg/mL. Prior to cryo-fixation, the vials were heat at the given temperature next to the cryofixation device using the dry block heating device and plunged into liquid ethane as fast as possible. All experiments were run during cooling from 130°C to RT. For technical reasons, the temperature during cryofixation could not be controlled. For this reason, the temperature associated with each cryo-TEM image in the text was related to the equilibration temperature (1 h) just before cryofixation, the device being as close as possible to the dry bath used for temperature control. Since temperature control during cryofixation was not possible in our device, and practically impossible at high temperatures, one should take the *T*\* values as indicative of the given value. Although the sample transfer between the dry bath and the cryofixation device was performed as fast as possible, it is more than likely that the actual temperature of the sample at the moment of cryofixation was lower than the one indicated here, although impossible to know precisely.

*Fluorescence microscopy.* Images were recorded using a 40x objective on a Leica SP5 upright confocal microscope with 405 nm laser excitation. The sample was marked with 6-dodecanoyl-N,N-dimethyl-2-naphthylamine (Laurdan, Mw= 353.54 g.mol$^{-1}$) fluorophore as follows. A stock solution of 3 mg of Laurdan in 100 µL of acetone was initially prepared (C= 30 mg/mL). A volume of 0.28 µL of the stock Laurdan solution was diluted in 1 mL of a 5 mg/mL G-C18:1-OH water solution temporary set at 110°C in a sealed Eppendorf. The latter was important so to enhance inclusion of Laurdan in the membrane constituted by the glucolipid, while in its vesicle phase. The final Laurdan concentration was approximately 24 µmol, with a molar Laurdan-to-G-C18:1-OH molar ratio of approximately 1-to-500. Fluorophore-to-lipid ratios above 1:200 were generally considered as non-interfering with the lipid phase behavior, as also verified in this work.

Laurdan can be excited between 365 nm and 410 nm and its emission wavelength can vary according to the rigidity of the membrane: intercalation within a rigid or flexible membrane generates emissions at 440 nm or 490 nm, respectively.[50] In the present experiment, we have flame-sealed a solution of G-C18:1-OH at 5 mg/mL containing Laurdan (1-to-500 in lipid-to-Laurdan ratio) in flat optical capillaries of 0.1 mm thickness. Capillaries were set on a Linkam support for temperature control, eventually placed under the microscope. Experiments were performed during the cooling cycle. The sample was excited at 405 nm and detection was made in spectral mode with a 5 nm window sliding every 5 nm from 412 nm to 602 nm in 16 bits, which allows quantify the signal and compare the different conditions. Quantitative analysis was performed according to the literature:[51,52] several ROI (range of interest) have been drawn on several images using the ROI manager of Fiji (is just ImageJ) software. The integrated intensity of each ROI at 440 nm and 490 nm was eventually recorded and used to calculate the value of the generalized polarization, GP, as follows of

$$GP = \frac{(\lambda_{440} - \lambda_{490})}{\lambda_{440} + \lambda_{490}}$$

A negative GP indicates a flexible membrane while a positive GP indicates a rigid membrane.

*Solid-state Nuclear Magnetic Resonance (ssNMR):* $^1H$ and $^{13}C$ ssNMR experiments were performed using an Avance III HD Bruker 7.05 T ($\nu_{1H}$= 300 MHz) spectrometer and a 4 mm magic angle spinning (MAS) probe. Temperature was controlled through a Smart cooler BCUII/a BCU-xtreme unit, using between 10 min and 20 min of equilibration time after each temperature variation. The exact equilibration time, which may vary for each temperature, is determined when the signal of the $^1H$ spectrum is stable over at least 5 minutes. Experiments were always performed during cooling from 110°C to 70°C. Experiments are run on a 50 wt% G-C18:1-OH sample prepared in $D_2O$ and placed in a disposable 30 µL Kel-F® insert, placed in a 4 mm zirconia rotor spinning at magic angle spinning, MAS= 2 kHz and chemical shifts were calibrated with respect to adamantane ($^{13}C$: $\delta_{13C}(CH_2)$= 38.48 ppm; $\delta_{1H}$= 1.91 ppm).

$^1H$ single-pulse experiments: number of transients, NS= 8; time domain size, TD= 16 k; pulse length, p($^1H$)= 2.06 µs; relaxation delay, D= 3 s.

2D $^1H$-$^1H$ spin diffusion (NOESY) MAS experiment: number of transients, NS= 96; time domain size, TD= 4 k; time domain size in indirect dimension, TD1= 512; incremental delay in indirect dimension, in= 390 µs; pulse length, p($^1H$)= 2.88 µs; relaxation delay, D= 1.2 s; mixing time, $t_{mix}$= 50, 100 and 300 ms.

Single pulse high-power decoupling $^{13}C\{^1H\}$ MAS experiments under high-power $^1H$ decoupling: number of transients, NS= 256; time domain size, TD= 2 k; pulse length, p($^{13}C$)= 3.4 µs; relaxation delay, D= 4 s; $^1H$ power level for high-power continuous wave decoupling, pl($^1H$)= 74.989 W.

**Results and discussion**

**Topologically-connected nanotubes and vesicles**

G-C18:1-OH is a biobased glucolipid, that is water-insoluble at room temperature, and whose physicochemical properties are poorly known. The DSC thermogram performed at 10°C/min of G-C18:1-OH for different water content (Figure 2a) shows two endothermic peaks on the dry powder (99.6 wt% glucolipid). They are attributed to two melting phenomena occurring at the following temperatures, $T_m$: 48.3°C ($T_{m1}$) and 95.7°C ($T_{m2}$), the enthalpy of the former being lower than that of the latter. Interestingly, DSC also shows that $T_{m1}$ is hydration independent and reversible (the associated exothermic peak is systematically observed), while the transition at $T_{m2}$ is simply suppressed (here tested up to 130°C) on hydrated powders and never observed on the cooling profile, even at slow cooling rates (1°C/min). The thermogram recorded on the dry powder is typical of single-molecule lipids having an intermediate,[53] metastable, phase (often referred to as the ripple phase, $P_{\beta'}$, in reference to the periodic undulation of the bilayer),[53–57] between the classical lamellar gel ($L_\beta$) and liquid crystalline ($L_\alpha$) phases[53–57] and characterized by regions of both order and disorder.[57,58] Water does not influence the low-temperature crystallization at $T_{m1}$, but it contributes to disrupt the ordered array of hydrocarbon chains in the metastable phase.

On the basis of the DSC thermogram, the aqueous phase behavior of G-C18:1-OH under diluted conditions (5 mg/mL) was studied at four different temperatures (red marks on Figure 2a): below and above $T_{m1}$, but also between $T_{m1}$ and $T_{m2}$, according to the thermogram of the dry powder.

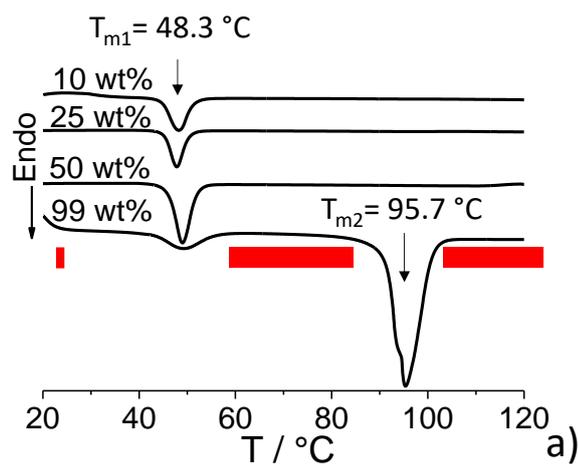

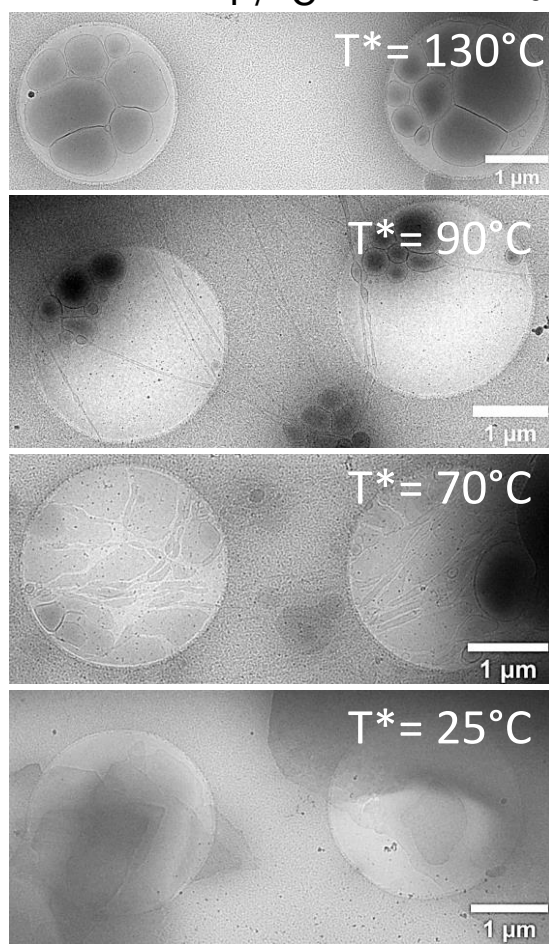

**Figure 2** – a) DSC thermogram of G-C18:1-OH under dry (0.4 wt% water, estimated by thermogravimetric analysis) and hydrated conditions recorded at a heating rate of 10°C/min. Red marks indicate the temperatures at which cryo-TEM experiments were performed on the corresponding 5 mg/mL aqueous solutions. b) Cryo-TEM images recorded for a 5 mg/mL G-C18:1-OH aqueous solution heat at $T^*$= 130°C, 90°C, 70°C and 25°C. $T^*$ indicates the temperature of the sample solution before cryofixation. $T^* \neq T$, T being the temperature of the sample upon plunging into liquid ethane. Please refer to the materials and methods section for more information.

The series of cryo-TEM images in Figure 2b shows an overview of the most relevant structures observed on G-C18:1-OH solution from $T^*=130°C$ to $T^*=25°C$, where $T^*$ is indicative of the equilibrium temperature before cryofixation (refer to materials and methods section for more information on the interpretation of $T^*$). At $T^*>100°C$, the sample is essentially composed of polydisperse single wall vesicles, massively shown by the complementary pictures presented in Figure S 2. In between $T_{m1}$ and $T_{m2}$, at $T^*=90°C$ and $T^*=70°C$ (Figure 2b, Figure 3, Figure 4, complemented by Figure S 3 and Figure S 4), vesicles are connected by nanotubes. At $T^*<T_{m1}$, flat crystals are observed (Figure 2b and Figure S 5).

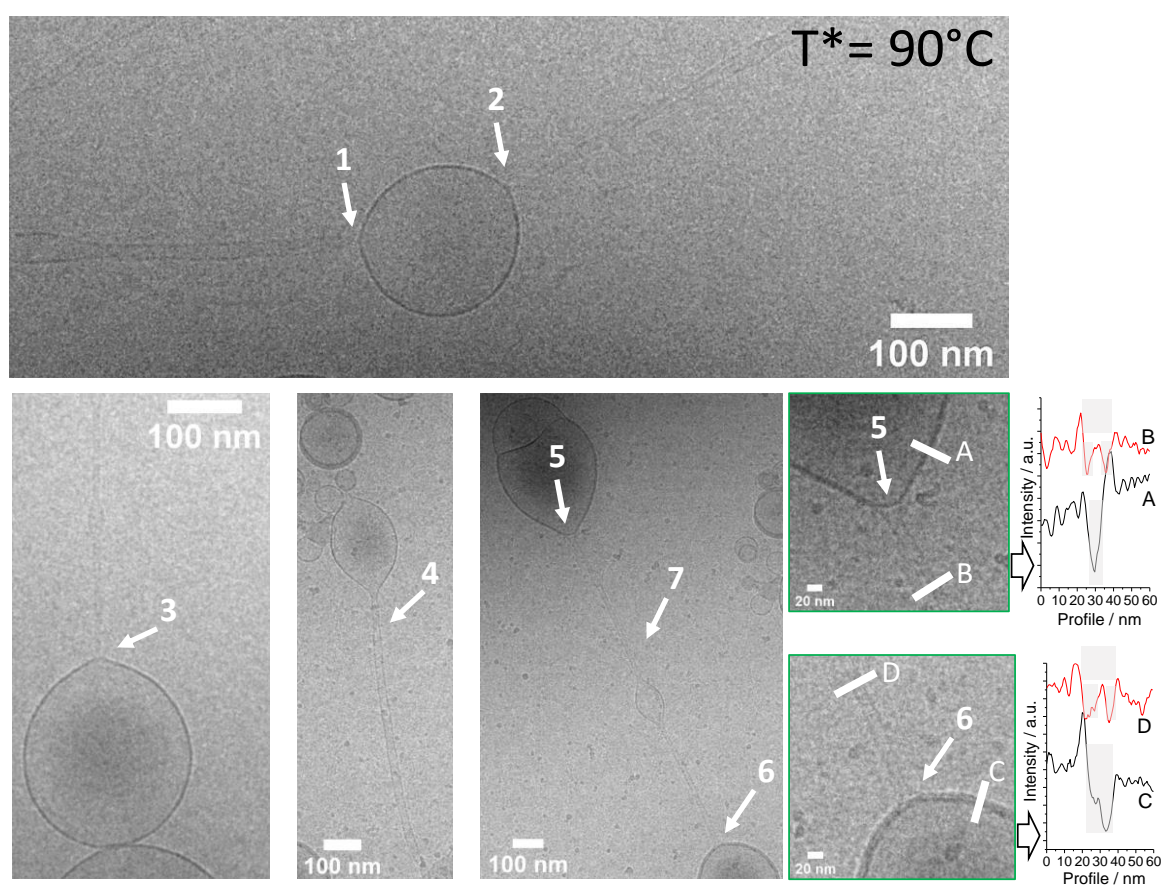

**Figure 3 – Cryo-TEM images recorded for a 5 mg/mL G-C18:1-OH aqueous solution heat at $T^*=90°C$. Arrows 1-6 points at topological continuum between vesicles and nanotubes. Arrow 7 indicates a typical nanotube. Letters A through D identify membrane regions of calculated thickness profiles.**

A deeper insight at $T^*=90°C$ is given in Figure 3 (more images are provided in Figure S 3). Arrows 1 through 6 mark very clearly the typical continuity in the glucolipid membrane between the vesicle and nanotubes. Arrow N° 7 shows that a single nanotube connects the outer membrane of two vesicles several hundred nanometers apart. The connection is shown by arrows N°5 and N°6. The corresponding profiles, A through D, confirm the tubular nature of the elongated objects. Profiles B and D are both characterized by typical symmetrical thick walls (low intensity depths highlighted by shaded rectangles) surrounding a hollow region (high intensity peak, between the shaded rectangles). The wall thickness varies between 3.5 and 4.0 nm for profile B and between 4.1 and 6.1 nm for profile D, while the total nanotube diameter varies between 14 nm at B and 20 nm at D. At the same time, the thickness of the vesicle wall is about 6 nm at A and 13.5 nm at C. Considerations about the bilayer or interdigitated structure of the membrane will be presented below.

The connection between nanotubes and vesicles seems to be multiple. For instance, arrow N°1 (Figure S 3) and N°4 (Figure 3) show a single-wall vesicle, of which the membrane is extruded into a nanotube. Similarly, a closer look at arrows N°1, 2 and 3 in Figure 3 show that the vesicle is single-wall and fully closed, while the nanotube nucleates at the vesicle outer surface, with no apparent topological continuity. In many regions of the sample at $T^*= 90°C$, one can find exvaginations or buds[18,20,34] in the nanotube wall. This is shown, for instance, by arrows N°7 and particularly put in evidence in Figure 4 by arrows N°1 through N°4, where the amplitude of the oscillation in diameter varies between 40-50 nm (arrow N°1) and 12-16 nm (arrows N°3-4). Exvaginations are membrane instabilities of Rayleigh-Plateau-type found both in simple (lipid bilayers)[21] or complex (organs)[59] biological matter. Buds are driven by a difference between the internal and external pressure,[30] but, in the absence of external (e.g., light) or internal (e.g., osmotic pressure) forces, the driving force can either be of biochemical origin in living organisms (e.g., disorders in ion pumps)[59] or ascribed to fluctuation of matter inside the tube in the case of synthetic lipid membranes.[21]

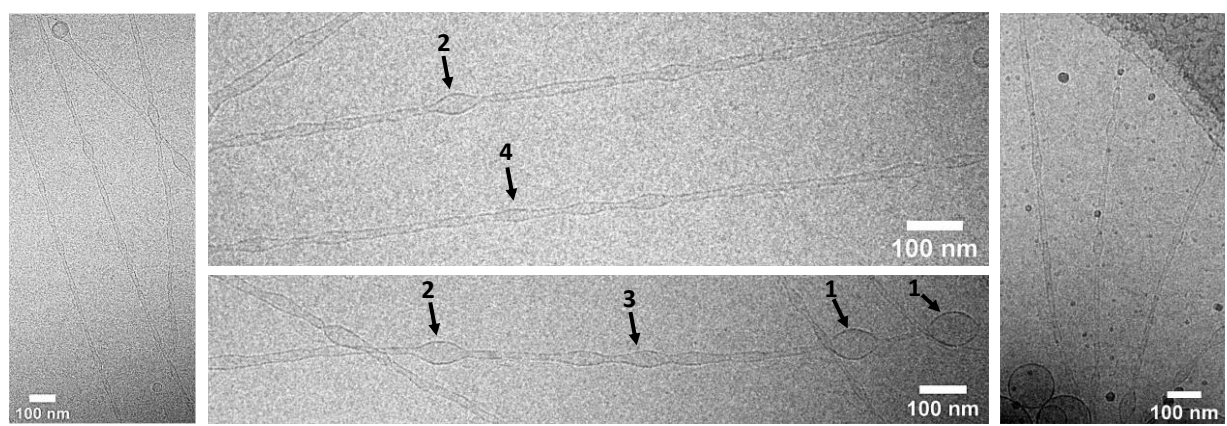

**Figure 4 - Cryo-TEM images recorded for a 5 mg/mL G-C18:1-OH aqueous solution heat at $T^*= 90°C$. Arrows 1-4 points at exvaginations observed in nanotubes.**

Still between $T_{m2}$ and $T_{m1}$, but at lower temperature ($T^*= 70°C$), the nanotubes seem to vary from single threads (arrow N°1, Figure S 4) to nanotubes of diameter below 15 nm, both connecting nanotubes of diameter ranging between 50 nm and 100 nm (arrows N°2, Figure S 4). The larger tubes are most likely formed by the inflation and flattening of the nanotubes, probably after assimilation of surrounding vesicles and fusion of the exvagination discussed above. The transition between vesicles, tubes and flat membranes is certainly accentuated. Arrow N°3 in Figure S 4 shows distinct vesicles while arrow N°4 identifies a flat structure evidently formed from fused vesicles. In particular, arrow N°5 points at the fusion of two vesicles while arrow N°6 points at two vesicles fusing into the flat membrane. Additional areas where vesicle fusion is prominent are indicated by the oscillations at the outer skirt of broad lamellae and pointed by arrow N°7.

Finally, below $T_{m1}$, at $T^*= 25°C$ for instance, mainly flat crystals, with no visible exvaginations or tubes, are visible (Figure S 5) by cryo-TEM and the material turns into an insoluble precipitate. Massive presence of flat crystalline structures is also indicated by the bulk iridescent behavior of a G-C18:1-OH solution under manual shear.

In summary, despite the uncertainty related to the actual state of equilibrium of the glucolipids at the moment of cryofixation, cryo-TEM depicts a vesicle phase at $T> 100°C$, a flat lamellar phase below $T_{m1}$ and an interconnected vesicle-nanotube network between $T_{m1}$ and about 100°C. The combination of statistically-relevant fluorescence microscopy and scattering/diffraction experiments provides an additional proof of the coexistence of vesicle/nanotube systems between the two transition temperatures. Please note that the experiments reported below were performed under controlled conditions of temperature, which is then indicated by the use of $T$, instead of $T^*$.

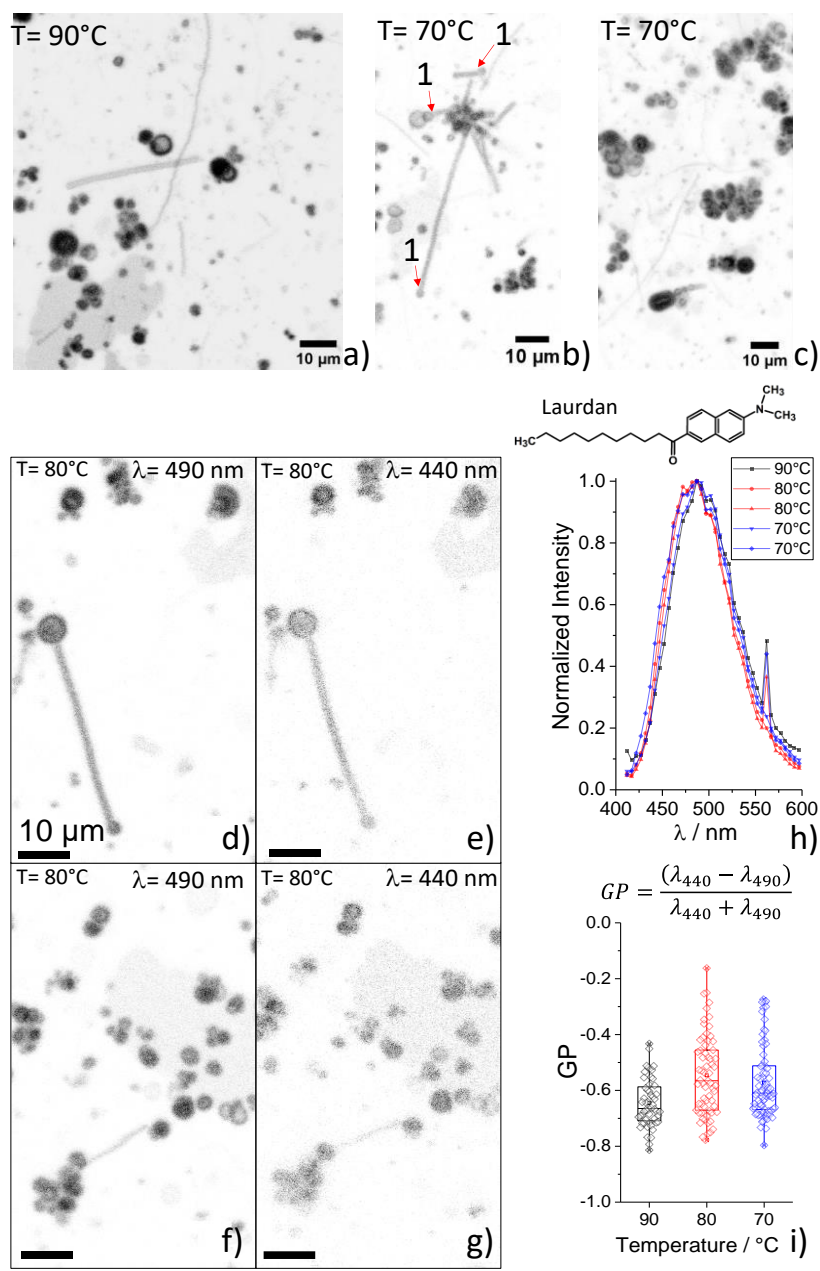

**Figure 5** – a-c) Confocal fluorescence microscopy images recorded at 90°C and 70°C for a 5 mg/mL G-C18:1-OH aqueous solution. d-g) Typical fluorescence microscopy images recorded at T= 80°C using Laurdan (refer to materials and methods section for more details on this fluorophore). Excitation wavelength is 405 nm for all the images, while emission wavelength is 490 nm in (d,f) and 440 nm in (e,g). h) Full emission spectra obtained from the integration of various fluorescence images recorded from 412 nm to 600 nm (step: 5 nm) at 90°C, 80°C and 70°C. i) Box plot of generalized polarization, GP, defined as $GP = \frac{(\lambda_{440} - \lambda_{490})}{\lambda_{440} + \lambda_{490}}$ with $\lambda_{440}$ and $\lambda_{490}$ being the emission wavelengths at 440 nm and 490 nm, at 90°C, 80°C and 70°C. Each data point corresponds to a set of regions of interest (ROIs) (54 at 90°C, 65 at 80°C and 75 at 70°C) selected in the corresponding fluorescence images. ROIs were representatively selected so to include the entire image but also vesicles alone, tubes alone and mixtures of vesicles and tubes.

Figure 5a,b shows two typical confocal fluorescence microscopy images recorded on a G-C18:1-OH solution (5 mg/mL) in a flame-sealed capillary heat at $T$= 90°C and $T$= 70°C. Although the resolution of optical microscopy is not comparable with cryo-TEM, the sample is broadly composed of vesicles, coexisting with nanotubes, whereas in some cases the nanotube nucleates from the vesicle itself (arrow N°1). Due to the resolution of confocal microscopy, vesicles of size in the order of 10 μm are more easily observed, although a wide range of smaller vesicles in the order of 1 μm and below are also detected. Similar considerations are also valid for the tubes.

From a qualitative point of view, topological connections between nanotubes and vesicles are observed both with confocal microscopy and cryo-TEM, which agree quite well. However, from a quantitative point of view, fluorescence microscopy experiments sometimes show tube diamters larger in size. According to theoretical arguments,[30] this could be explained by a non-equilibrated system. Dimova et al.[30] have shown that when membrane composed of cylindrical (DOPC) and conical (DOPE-PEG or $G_{M1}$ ganglioside) phospholipids deform into a cap-ended nanotube, the diameter of the tube at equilibrium is unique and quantifiable. However, if the idea that the nanotube diameter is somewhat related to the tube's bending modulus may not unreasonable, that approach was developed for pulling membrane tubes and based on the hypothesis that vesicles and nanotubes have the same bending modulus and, consequently, the same chemical composition. These assumptions are in contrast with the extensive literature on lipid nanotubes,[24,35,36] the latter being rich in conical lipids and known to have higher bending energy than vesicle membranes.[8]

In the present work, cryo-TEM images were recorded on samples thermalized during about one day while fluorescence microscopy were recorded on samples thermalized in the order of the hour. Considering that the solid-state NMR $^1$H signal of G-C18:1-OH (NMR experiments are discussed further below) requires about 10 to 20 min to equilibrate after thermalization and that the SAXS signal does not evolve on the minute time scale upon cooling, it is reasonable to exclude, at a first approximation, a non-equilibrated process. However, being the equilibration time in similar systems known to be sometimes excessively long (days, weeks or even months),[27,60] a specific time-dependent study will certainly be welcome in the future.

If one retains the idea that the nanotube diameter is somewhat associated to its bending modulus, the variability in the diameter could only be associated to fluctuations in the structure of the tubes, their chemical composition (G-C18:1-OH only) being the same as in the vesicles. The structure of the tubes, and in particular the difference in the G-C18:1-OH arrangement in tubes and vesicles, will be discussed in detail later in this work. Finally, it should also be noted that the work of Dimova et al. was valid within the context of pulled nanotubes, while here one finds both pulled (Figure 3, arrows 4, 5 and 6) and nucleated (Figure 3, arrows 1 and 2) nanotubes.

**Structural study of vesicle-nanotube systems**

SANS recorded at $T$= 90°C (Figure 6a) shows a scattering profile with a pronounced -2 dependence of the scattering intensity against the wavevector q in the log-log plot. This indicates the massive presence of a flat interface and it is typical of bilayers in vesicular morphologies.[61] However, the typical signature of nanotubes cannot be observed in the SANS profile, and this can be explained by both an argument related to the relative lower amount of nanotubes with respect to vesicles at 90°C but also to the steeper slope of flat morphologies (-2) compared to tubes (-1), the signal of the former masking the tube signal at low wavevector values. The SANS profile is also characterized by a diffraction peak at $q$= 1.67 nm$^{-1}$, discussed in more detail below.

Diffraction experiments were performed on samples at various concentrations as a function of temperature. The SAXS profiles of the sample at the same low concentration (5 mg/mL) as the cryo-TEM and microscopy experiments show (Figure S 6) peaks indicative of lamellar order, of which the intense (001) peak varies between 1.58 nm$^{-1}$ at 25°C and 1.79 nm$^{-1}$ at 111°C, in agreement with SANS. Considering the low intensity of the (002) peak, additional experiments were run at higher concentration (50 wt%) and using a more performing temperature-resolved *in situ* protocol. Selected data in Figure 6b confirm the shift of the (001) reflection of the lamellar phase to lower q-values, while the full SAXS dataset in Figure S 7 shows a well-defined structural transition at about 48°C. This is attributed to the transition between the crystal and liquid crystalline phase, as also shown in the 2D contour WAXS plot (Figure S 7): the broad peak at 14.9 nm$^{-1}$ above 50°C is characteristics of the lateral packing of lipids in bilayers. While its position could be consistent with an L$_\beta$ phase,[62,63] its broadness and positional invariance as well as the coexistence between vesicles and tubes do not allow a precise assignment.

*In situ* SAXS shows significant structural differences during the cooling cycles, and in particular the appearance of an extra peak, (001)* (Figure 6b), at about 1.83 nm$^{-1}$. This peak starts to appears at 110°C as a broad shoulder to (001), it coexists in the region between 90°C and 70°C and it disappears at room temperature. Very interestingly, an extra broad peak is also observed above 4 nm$^{-1}$ in the SAXS profiles of sample at 5 mg/mL. Since *in situ* WAXS (Figure S 7) does not show any significant shift in the peak at 14.9 nm$^{-1}$, the structural study corroborates the hypothesis that two self-assembled forms of G-C18:1-OH coexist upon cooling. The structures may differ is their morphology and supramolecular long-range packing, although locally they are both characterized by the same liquid crystalline intra-membrane packing of the acyl chains.

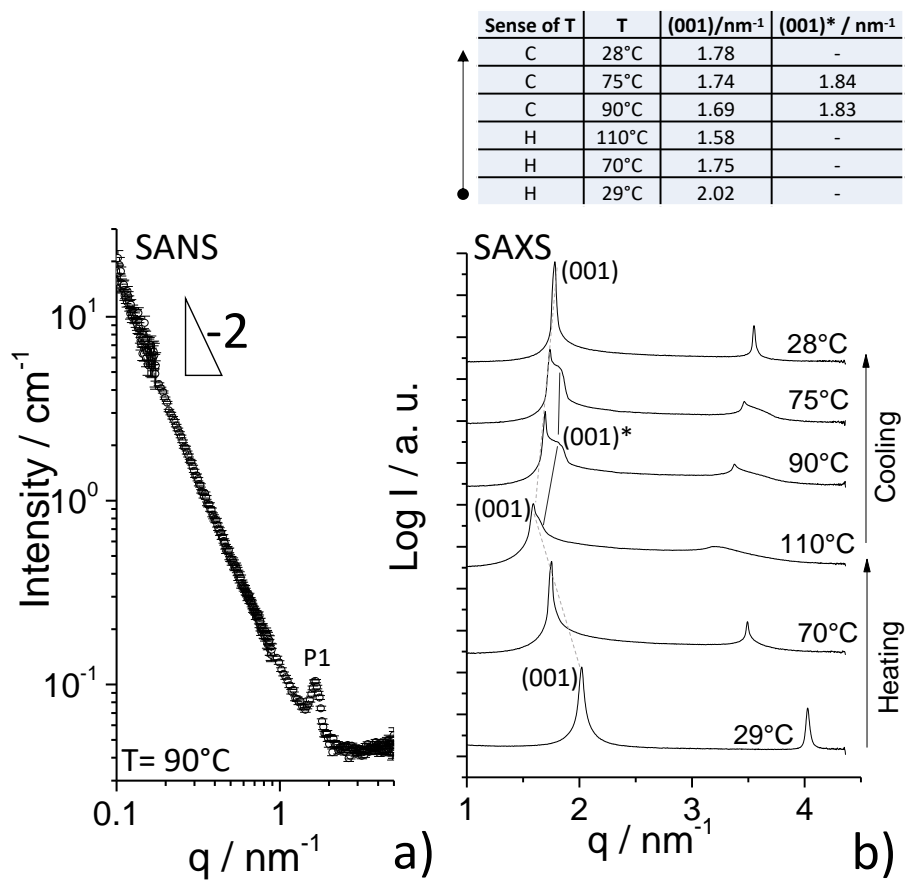

**Figure 6 – a) SANS experiment recorded at 90°C for a 5 mg/mL G-C18:1-OH aqueous (D$_2$O) solution. b) SAXS profiles extracted from the continuous heating-cooling temperature-resolved *in situ* SAXS contour plot in Figure S 7. Experiment is performed on a 50 wt% G-C18:1-OH aqueous solution. The peak position and sense of temperature (T) variation (H: heating; C: cooling) corresponding to the SAXS profiles is given on top of the plot in b).**

Does G-C18:1-OH assume a bilayer or an interdigitated structure, the latter being the one found for other similar bolaform glucolipids?[45,46] The size of G-C18:1-OH can be estimated to be at the most 3.2 nm, given by the length, L, of the C16 tail (L = 2.2 nm, from the Tanford formula, L= 1.54 + 1.265 × n, (n= 16)[64]) and the sugar (about 1.0 nm).[45,46] However, the *cis* conformation of the C=C bond imposes a "kink" of 60° in the aliphatic chain, which reduces the effective length of the oleic backbone to about 1.7 nm, for an effective total G-C18:1-OH length of about 2.7 nm. Cryo-TEM provides a tube thickness varying from 3.5 nm to 6.0 nm and a vesicle membrane thickness of at least 6.0 nm. WAXS suggests an interlamellar thickness (membrane and water) between 4.0 nm and 3.6 nm and an inter-lipid distance in the tube of about 1.6 nm. In the absence of a clear form factor oscillation profile, as found for similar glucolipids,[45,46] one must acknowledge that these data indicate neither a fully interdigitated nor a double layer structure and might suggest a tilted lipid arrangement with partial interdigitation.[65]

**G-C18:1-OH can assume two coexisting, head-head and head-tail, configurations**

Vesicles are characterized by low positive mean and Gaussian curvatures, while nanotubes are described by a combination of a high positive mean and zero Gaussian curvatures. In this regard, the coexistence of vesicles and nanotubes requires glucolipids to adapt themselves to both environments at the same time. For this reason, the formation of nanotube-vesicle networks generally requires a specific driving force, able to impose a high mean and zero Gaussian curvature to the membrane. In many systems, insertion of at least a second constituent (lipid, surfactant, protein) in the membrane becomes a necessary internal driving force for the nucleation and growth of so-called membrane inclusions, from which topologically-connected nanotubes can form,[10–12,35,66] as explained by orientational ordering arguments.[67] Combination of membrane formulations with additional internal driving forces (e.g., polymer-polymer phase separation)[30] or external driving forces, such as use of laser "tweezers",[33] osmotic pressure,[6] electroformation[5–8,31,32] or controlled suction,[13–16] are otherwise necessary, as also supported by theoretical studies.[23,30]

The bending energy of tubes is higher than that of a vesicle membrane,[8] due to the dependence of the membrane free energy from the principal curvatures. For this reason, isotropic (rotational symmetry along the main axis), but also anisotropic,[36] lipids generally stabilize low-curvature shapes, like flat membranes or vesicles. Lipid nanotubes are also a common phase formed by amphiphiles, but generally for lipids with a certain complexity in their molecular structures, as exampled by diacetylenic derivatives of phospholipids, among others and always close to, and generally below, their $T_m$.[28,29,68–70] Negative contributions to the bending energy favoring the simultaneous presence of interconnected vesicles and tubes was shown to be driven by the presence of anisotropic lipids in a membrane continuum. Spontaneous deviations to the elastic theory in the presence of surfactants, lipids and proteins with anisotropic structures favor the formation of protrusions with radius of the order of the fraction of a micron.[9–12,18,23–26,67]

In the absence of both internal and external driving forces mentioned above, G-C18:1-OH should behave as a standard lipid. Its DSC profile (Figure 2)[56,71,72] under both hydrated and dry conditions identifies a hydration-independent transition ($T_{m1}$) slightly above room temperature; this is followed by one main transition at higher temperature ($T_{m2}$) in the dry sample.[53] For such a profile, one expects a gel phase ($L_\beta$) below $T_{m1}$ and a liquid crystalline structure, possibly in the morphology of vesicles ($L_4$ phase), possibly above $T_{m1}$, and most likely above $T_{m2}$,[34] as experimentally found. On the basis of the thermogram recorded on the dry powder, one may also expect a "ripple" ($P_\beta$) phase between $T_{m1}$ and $T_{m2}$ characterized by periodic undulations of the bilayer (generally addressed to as regions with both liquid order and disorder).[57,58]

In the present work, instead of a $L_\beta$ phase, we observe flat crystals below $T_{m1}$, and a vesicle-nanotube network instead of the "ripple" phase. The latter is particularly unexpected because the membrane is composed of a single molecule (G-C18:1-OH) and no anisotropic external force (osmotic, hydrodynamic, electric) is used to pull and stabilize the tubes. In addition, the nanotubes are observed at fairly high temperatures, above $T_{m1}$ (between about 60°C and 90°C), while this morphology is generally favored at low temperature (proportionality with 1/kT, figure 5 in Ref. [67]), commonly below a given transition temperature, characteristics for each specific amphiphile, lipid or block copolymer.[27,29] Furthermore, nanotubes of G-C18:1-OH seem to be stable in time, thus differing from what is found in electroformed vesicles.[7,8] They are also observed away from phase transition events (may them be at $T_{m1}$ or $T_{m2}$), as otherwise reported before for a diacetylenic derivative of phosphatidylcholine[28] and many other systems.[29]

A stable nanotube-vesicle network for a single-molecule phospholipid in the absence of controlled external driving forces and above its $T_m$ is not expected and has never been reported before. It could be explained by the coexistence of two micro-separated phases of different rigidity. This hypothesis is excluded by confocal microscopy on Laurdan-labelled membranes and magic angle spinning (MAS) solid-state NMR (ssNMR) experiments.

Laurdan is a well-known fluorophore used to put in evidence the coexistence of rigid and fluid domains in both model and biological membranes.[50–52] Laurdan was used here to label the G-C18:1-OH membranes and to verify the possible micro-segregation and coexistence of rigid (e.g., $L_\beta$, $\lambda_{em}$= 440 nm) and fluid (e.g., $L_\alpha$, $\lambda_{em}$= 490 nm) domains, in tubes and vesicles, respectively. A set of quantitative experiments performed on a 5 mg/mL solution, summarized in Figure 5, shows an excess of fluorescence at only $\lambda_{em}$= 490 nm for temperature between 90°C and 70°C (Figure 5h). The GP value, which quantifies the rigid/fluid character of the membrane, is systematically negative (Figure 5i), indicating an overall fluid membrane at all temperatures. GP is quantified on more than 50 regions of interest (ROIs) at each temperature, including vesicles and tubes alone as well as connected nanotube-vesicles (Figure 5d-g). The dispersion of GP at 80°C and 70°C between -0.8 and -0.2 could suggest the spurious presence of membranes with various degrees of flexibility, although fluidity is always predominant.

The combination of WAXS (broad peak at 14.9 nm$^{-1}$, invariant with temperature, Figure S 7) and confocal fluorescence microscopy using Laurdan confirms the single-phase of the membrane composed of G-C18:1-OH, both in vesicles and nanotubes. This is also consistent with the lack of phase transitions in DSC experiments between ~85°C and ~60°C, especially in hydrated samples. We conclude that the vesicle-nanotube transition is not driven by the local segregation of gel phase (L$_\beta$) microdomains, nor that the nanotubes are composed of such phase.

ssNMR is the best-suited approach to investigate the molecular origin for the nucleation of anisotropic inclusions, of which the average orientation depends on the local membrane and inclusion curvatures and was shown to promote tubing.[67] ssNMR can provide information on the local molecular mobility, e.g. through the nuclear spin relaxation behavior, but also on intermolecular interactions and stereochemistry, e.g. through variations in the chemical shift.[73–75]

Ideally, ssNMR should be performed on diluted G-C18:1-OH samples above $T_{m1}$. However, the rates of magic angle spinning (MAS), generally above the kHz and necessary for high resolution, are too high for low-concentrated samples, which are centrifuged on the sample holder walls. ssNMR must then been performed on concentrated samples, here chosen at 50 wt%, as classically done for phospholipids.[76] Figure 7a-c shows the evolution of the chemical shift position of each $^{13}$C peak of G-C18:1-OH with temperature, during cooling from 110°C to 30°C (selected full spectra are given in Figure S 8). The chemical shifts are given relative (%) to the shift at T= 110°C. All signals of the sugar headgroup become more shielded (smaller values), between -0.2% for C4' and -1% for C6', while the signals related to the aliphatic chain have a double behaviour. The carbon sites at the extremes (C1, C2, C16) and around the double bond (C8, C9, C10, C11) become more shielded, with the extreme (C1, C16) being more affected (between -0.7 % and -1%). All other carbons in the aliphatic chain (C3, C14, C15, -CH$_2$-) become deshielded (higher values), with variation up to +1.2% for the intra-chain –CH$_2$- sites.

Strong variations in the chemical shift of carbon atoms in the solid-state is a typical sign of conformational rearrangements. In the specific case of glycolipids, specific self-assembled structures (monolayer crystal, bilayer crystal or micellar fiber) have been associated to conformational changes in the linear gluconamide headgroup by mean of $^{13}$C ssNMR.[77–80] It was found that *gauche* effects, in some cases up to the γ position, in the sugar chain induce upfield chemical shifts, while *anti* and *trans* conformations result in downfield shifts. Each structure (monolayer crystal, bilayer crystal or micellar fiber) was then characterized by a given organization of the gluconamide and a characteristic $^{13}$C ssNMR fingerprint, which could be explained by *gauche*, *γ-gauche*, *anti* and *trans* conformations of the headgroup.[77–80] Similar structural-spectroscopic correlations have been collected over the years for much simpler cyclic sugars[81] including glucose,[82–84] whereas pyranoses can undergo important conformational changes when the appropriate energy barriers are crossed.[85–88]

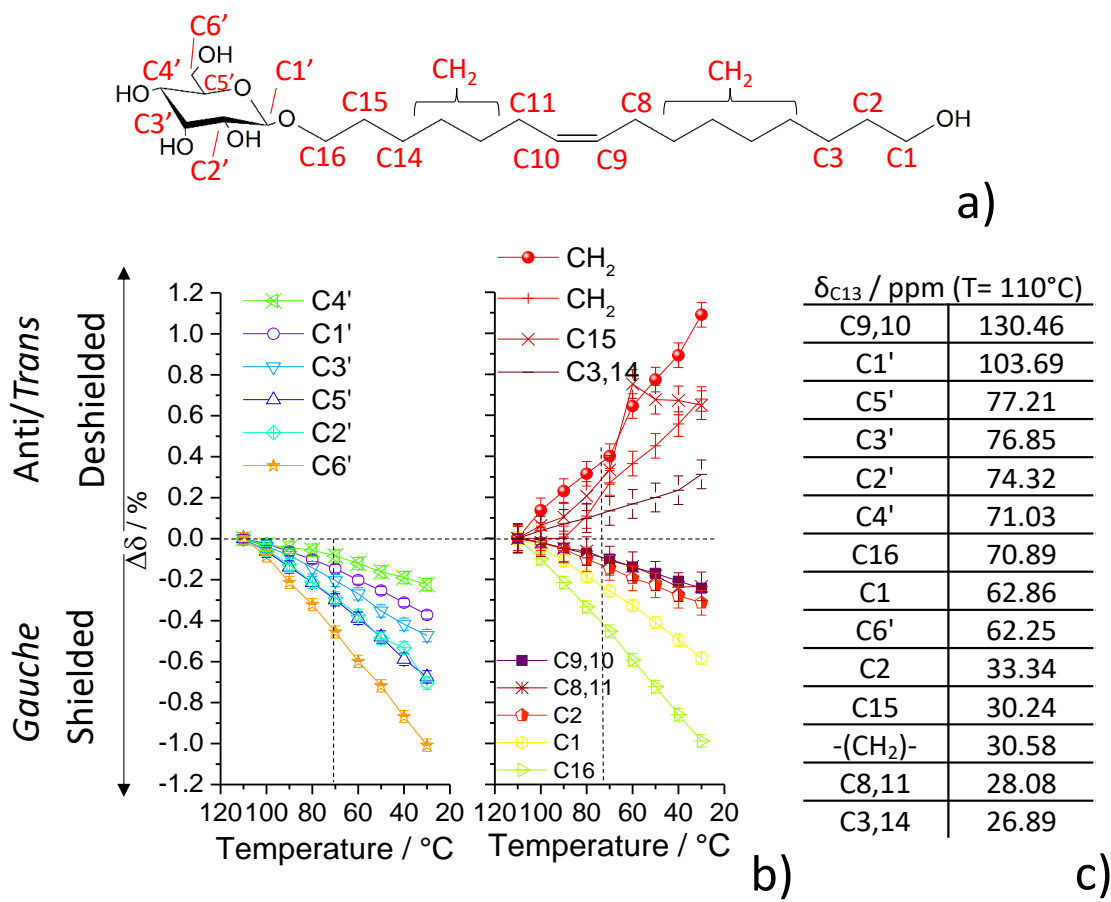

**Figure 7 – Temperature-dependent evolution of the $^{13}$C chemical shift of a 50 wt% solution of G-C18:1-OH: a) labelling, b) relative shift given in % with respect to the chemical shift measured at T= 110°C, c) list of chemical shift values at T= 110°C. Error bars are calculated by propagation of the uncertainty in reading the chemical shift value (±0.01 ppm). Experiments are recorded using a high-power decoupling $^{13}$C{$^{1}$H} single-pulse MAS sequence. Typical $^{13}$C spectra at 110°C and 70°C are given in Figure S 8.**

A quantitative conformational study based on the $^{13}$C chemical shift variations shown in Figure 7b would be certainly helpful, but very delicate in the context of the present state of the art, because a rigorous approach requires at least one known crystal structure,[77,78] which is currently missing for this compound. From a qualitative point of view, shielding demonstrates important *gauche* conformational effects[77–80,82–84] of the glucose, but also in the close vicinity of the OH end-group. Deshielding demonstrates *anti* and *trans* effects in the middle of the aliphatic tail, far from the C=C bond and extremes. This is actually expected for an alkyl chain.[89] These effects are important below 90°C, precisely in the region of coexistence between vesicles and nanotubes. One could actually expect the splitting of selected $^{13}$C signals, demonstrating two chemically-inequivalent environments, as reported for the *cis* olefin group in various phases of oleic acid in the vicinity of 0°C.[90] However, the time-scale of the NMR experiment is most likely too long compared to the diffusion of G-C18:1-OH between the vesicle and nanotube structures, thus resulting in an average $^{13}$C signal.

According to NMR, cooling below 100°C induces important and continuous variations in the conformations of glucose and fatty alcohol moieties of G-C18:1-OH. These occur within a single-phase domain (as deduced by WAXS) of equivalent membrane rigidity (as derived from Laurdan probe in confocal microscopy). Although never reported before, differences in the conformation of the same amphiphile (here, G-C18:1-OH) could satisfy the theoretically-required conditions, that induce local lipid segregation and formation of membrane inclusions (Figure 8b). In case of their non-zero average orientation, the local membrane curvature increases,[67] and nanotubes can spontaneously form (Figure 8). More intriguing, this mechanism occurs up to at least 10°C above $T_{m1}$, although it should not be favored at all above the $T_m$.[27,29]

From both experimental and theoretical backgrounds, the conditions of topologically-connected vesicles-nanotubes systems are met only when either internal (two molecules and/or polymer-polymer phase separation), or external, driving forces are applied.[23,30] In the present single-lipid system, of which the phase behavior is summarized in Figure 8a, the most plausible driving force could be explained by the NMR results. G-C18:1-OH could occupy two molecular environments, characterized by two distinct conformations of the glucose and the acyl chain.

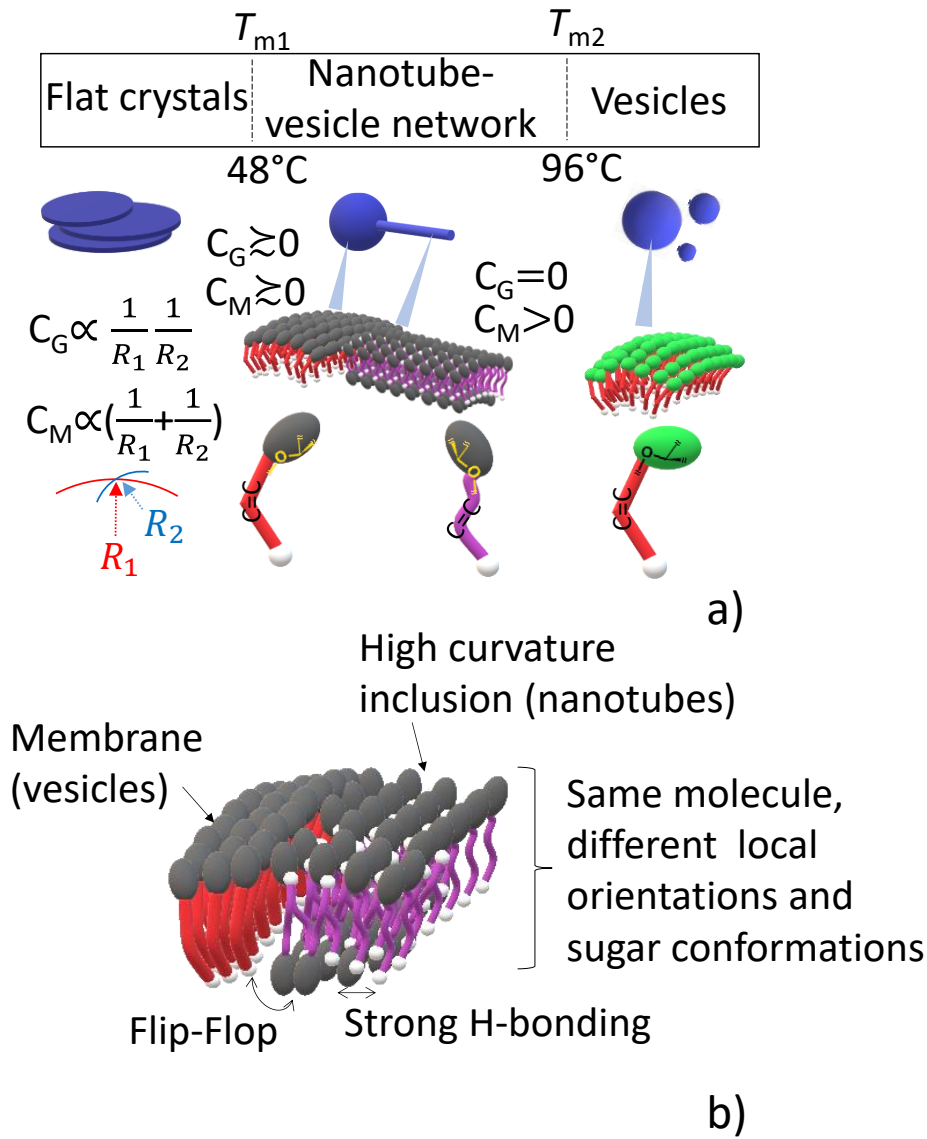

**Figure 8** – a) Thermotropic phase behavior of G-C18:1-OH at 5 mg/mL. Note that $T_{m2}$ is observed on the dry powder, only. Possible relationship between the molecular environment and membrane curvatures. All the molecular drawings refer to the same G-C18:1-OH molecule (bulky ellipsoid ≡ sugar headgroup), while each color corresponds to a given conformation. The curvature, C, is defined as 1/R, R being the curvature radius. $R_1$, $R_2$: radius of curvature; $C_G$: Gaussian curvature; $C_M$: Mean curvature. b) Free interpretation of the anisotropic, high-curvature ($C_M > 0$; $C_G = 0$), inclusions in the membrane (low $C_M$ and $C_G$) continuum driven by inter-sugar H-bonding and reduced intramembrane diffusion for a G-C18:1-OH glucolipid.

Two related questions are unanswered at the moment. What actually drives nanotube formation and is the organization of G-C18:1-OH within the vesicles and tubes the same? Lipid nanotubes have been studied for at least three decades, over which broad experimental and theoretical insight have been generated. Despite this fact, the actual understanding is still matter of debate.[29] Micelle-to-nanotube and vesicle-to-nanotube phase transitions are generally reported for single amphiphiles below a given $T_m$ (or $T_g$ in the case of block copolymers[27]) and intermediate phase composed of twisted and flat fibers is concomitantly reported. Initially thought to be a chirality-driven phenomenon, recent studies propose a more general mechanism involving crystallization and symmetry breaking in the lipid packing driven by optimal packing.[29] If twisted ribbons were observed before for a number of microbial glycolipid bolaamphiphiles,[47] they are not found for G-C18:1-OH. This fact, as well as the large body of cryo-TEM images and the observation of nanotubes above $T_{m1}$, rather indicate a direct topological change from positive (vesicles) to zero (nanotubes) Gaussian curvatures. In this case, vesicle-to-nanotube transitions could be driven by external forces.[5–8,13–16,31–33]

The intrinsic experimental design of the present system excludes both controlled external (e.g., laser tweezers or osmotic pressure) and internal (e.g., polymer-polymer phase separation) forces. For this reason, the possible explanations are limited to two: spontaneous budding of the membrane or nanotube pulling after inter-vesicle collisions due to Brownian motion, possibly combined to internal convection flow, reasonably expected at high temperature. Inclusions composed of a different arrangement of the same molecule in the membrane continuum can theoretically explain spontaneous nanotube formation in an isotropic medium.[6,7,9–12,18,22–26,67] Inter-vesicle collisions are probably more likely to prevail, because they could explain the large number of nanotubes connecting two distant vesicles. This hypothesis is also supported by the fact that nanotube-vesicle networks are mainly observed upon cooling from the vesicle phase above 100°C. Furthermore, cryo-TEM images (Figure 2a, Figure S 2, Figure S 3, arrow 2) show evidence of contacts between vesicles. Interestingly, similar events occur for any phospholipid-based vesicular solution, but nanotubes are extremely rare unless at least two lipids of different molecular structure are mixed together. The ssNMR data strongly suggest that the coexistence of two G-C18:1-OH configurations, involving both the sugar headgroup and fatty alcohol chain, are possible for this molecule. As a matter of fact, this is not so surprising, as more than one energy minimum, each corresponding to a given conformation, are known for glucose, its non-reducing derivative, methylglucose, and carbohydrates in general.[86–88]

The presence of two configurations of G-C18:1-OH could explain the coexistence between two structures of different curvature, but the $^{13}$C MAS NMR are ambiguous on this point and further experiments are required. G-C18:1-OH is a bolaform amphiphile and flip-flop could be a possible mechanism involved in the nanotube stabilization, whereas head-tail interactions have long been considered as important in nanotube-forming bolaform amphiphiles.[65,68] We address this question by exploring intermolecular interactions in a direct manner, using 2D $^1$H-$^1$H homonuclear ssNMR correlation spectroscopy employing spin diffusion (NOESY) experiments. Off-diagonal peaks in typical 2D contour plots demonstrate a proximity between two chemically-inequivalent $^1$H sites, as sketched in Figure 9a. The distance range between protons is controlled by mean of the spin diffusion mixing time, $t_m$. Low values of $t_m$ only probe very close neighbors (<< 5 Å) and generally located on the same molecule. Long $t_m$ probe long distances (>> 5 Å). Appropriate $t_m$ values can, on the contrary, probe medium-range (about 5 Å) distances, thus discriminating between molecular configurations in space. In the hypothesis of flip-flopped G-C18:1-OH (Figure 9c), one expects medium-range interaction between $H_1$ (R-C$\underline{H_2}$-OH) and the glucose headgroup ($H_{1'}$-$H_{6'}$), while these groups are separated by more than 2 nm, and should not interact in a parallel configuration (Figure 9b). Presently, $t_m$ is optimized between 50 ms and 300 ms (Figure S 9a-c and Figure S 10a-c) and the value of 100 ms (Figure 9b) is found to be the most appropriate one, at the limit between short- and medium-range distances.

Highlight of the glucose headgroup region in spin diffusion experiments at 110°C (Figure 9b) shows two off-diagonal cross-peaks (*x*-, *y*-labelled) attributed to intra-glucose interactions. At 70°C (Figure 9c), an additional cross-peak (*z*-labelled) identifies an interaction between $H_{1'}$ and $H_{4'/6'}$. Further proof of the flip-flopped configuration at 70°C is given by a deeper analysis of the entire set of $^1$H-$^1$H correlation plots provided in the Supporting Information corresponding to Figure S 9 and Figure S 10.

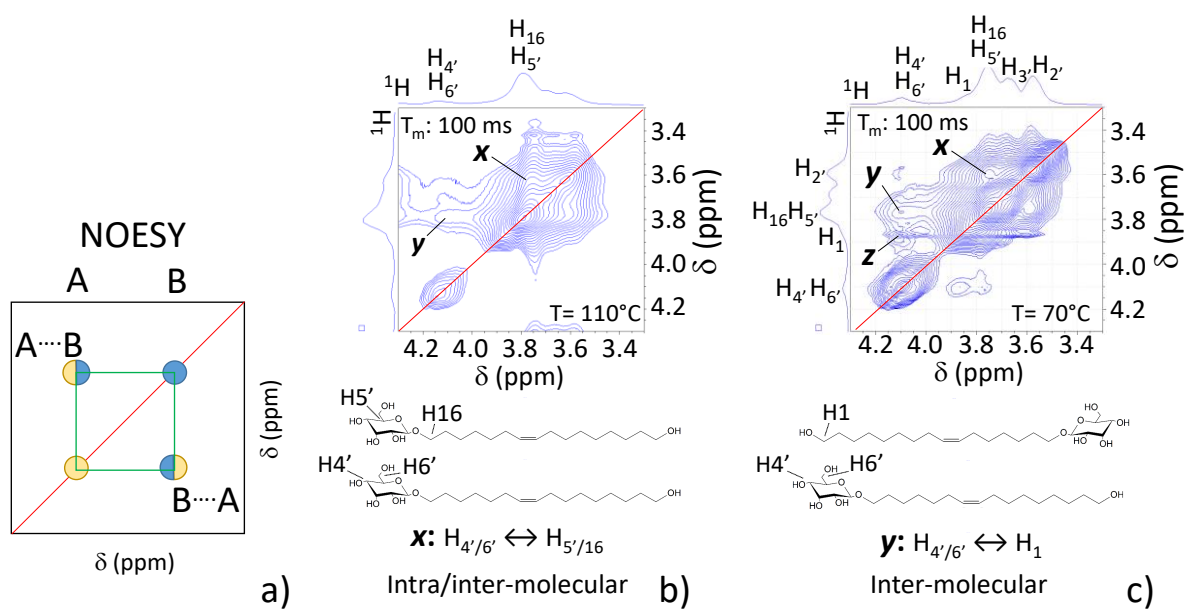

**Figure 9 - 2D $^1$H-$^1$H spin diffusion (NOESY) NMR experiments (glucose region) using 100 ms of mixing time recorded on a 50 wt% G-C18:1-OH sample in D$_2$O. Sample temperature is regulated to b) 110°C and**

c) 70°C, reached by cooling from 110°C. The typical scheme illustrating how to interpret the 2D experiments is reported in a). Full 2D contour plots recorded at mixing time of 50 ms, 100 ms and 300 ms are given in Figure S 9 and Figure S 10.

Homonuclear 2D ssNMR supports the idea, according to which G-C18:1-OH can undergo molecular flip-flops and a head-tail arrangement at 50 wt%. Extrapolating these conclusions in the lower concentration range, lack of proximity between $H_1$, $H_2$ and the sugar headgroup at T= 110°C suggests a head-head configuration in the vesicle-rich region (Figure 9b) and head-tail configuration (Figure 9c) within the nanotubes. It is however still unclear whether flip-flops are driven by spontaneous, temperature-induced, conformational changes in the glucolipid packing, stimulating inclusions within the membrane itself,[6,7,9–12,18,22–26,67] or by defects generated by uncontrolled inter-vesicle collisions due to the high temperature profiles. In fact, a combination of both is not unlikely. It is also unclear whether or not head-tail or head-head arrangement have the same energy and produce membranes with similar bending modulus. It is also unclear whether or not the relative content of head-tail or head-head arrangement is stable or it varies with temperature. However, flip-flops could actually induce local fluctuations in the surface tension of the membrane and in analogy with fluctuations in the membrane composition,[21] this could at least explain budding in the tubes (Figure 4). As explained previously in the manuscript, within the limits of the hypothesis formulated by Dimova et al.,[30] one could also explain the heterogeneity in the tube diameters by fluctuations in the tube's bending modulus, associated to variations in the relative content of head-head and head-tail arrangements. This hypothesis should be verified further.

In an energy landscape approach,[91] the head-head configuration is certainly favored at high temperature, while both head-head and head-tail configurations could be favored in the mid-temperature region, corresponding to the range between $T_{m1}$ and $T_{m2}$ in the dry powder. Both statistical and defect-induced flip-flop events contribute to form high mean, zero Gaussian, curvature morphologies (nanotubes), for which the head-tail conformation, limiting the proximity between bulky sugar headgroups, is probably more adapted (Figure 8b). The flip-flop interpretation now reasonably explains the strong differences in the sugar and fatty alcohol conformations, deduced by the multiple NMR signals attributed to chemically-equivalent groups. The existence of various low-energy conformations, each having specific NMR signature, for sugars in solution[85–88] and solid-state[77–80,92] further supports our interpretation.

**Conclusions**

Topologically-connected nanotube-vesicle lipidic systems are commonly observed for self-assembled membranes composed of at least two amphiphiles, and spontaneously driven by anisotropic inclusions and/or by external directional forces. This work shows that strong differences in membrane curvature can spontaneously occur in single-lipid systems without external anisotropic forces. Coexistence and interconnection between nanotubes and vesicles are evidenced via cryo-TEM and fluorescence confocal microscopy in a single-phase region at temperature above a transition temperature, $T_{m1}= 48.3°C$, determined by DSC on the hydrated sample. X-ray diffraction (WAXS) also shows the coexistence of two lattice periods, one attributed to an inter-membrane repeating distance and the other possibly related to the intra-nanotube glucolipid arrangement. Labelling the systems with Laurdan, a fluorophore of which the emission strongly varies with the local membrane rigidity, excludes the presence of rigid micro-separated phases (e.g., $L_\beta$).

$^1$H and $^{13}$C ssNMR confirm the presence of one main fluid G-C18:1-OH environment above $T_{m2}$. Below $T_{m2}$, the splitting of some $^{13}$C peaks and the coexistence of short and long $^1$H relaxation components demonstrate that G-C18:1-OH can simultaneously pack into structurally and dynamically inequivalent environments. Combining 2D $^1$H-$^1$H correlation experiments based on spin diffusion suggests that vesicles could contain a preferred head-head configuration of the bolaform lipid at 110°C, in the vesicle region of the phase diagram, while nanotubes a preferred head-tail assembly below 110°C.

The presence of two environments composed of the same molecule characterized by different molecular configurations and rigidity fulfills the theoretical conditions necessary to spontaneously nucleate a membrane inclusion of non-zero average orientation, described as necessary to observe topologically-connected nanotube-vesicle structures in the absence of external forces. This hypothesis could also be enriched by the possibility of nanotubes to be pulled away between two vesicles after a collision event, as suggested by cryo-TEM analysis.

On a general level, this work illustrates the unique properties of a new family of amphiphiles, of which the combination of bolaform structure and chemical groups (glucose, alcohol) merge in a single compound the functionality commonly reported for more complex formulated systems. More specifically, it expands the actual theoretical and experimental framework on lipid networks stabilized by a different radius of curvature. This work could also motivate the synthesis, or isolation, of new lipids providing a higher level of structural complexity and 3D architectural control, possibly in the range of room temperature, for the development of the next generation of soft materials.

**Supporting Information**

Supporting information files contains characterization of G-C18:1-OH (HPLC-ELSD, composition table, $^1$H NMR), additional cryo-TEM images recorded on aqueous solutions of G-C18:1-OH in water at T*= 130°C, 90°C, 70°C, 25°C, WAXS experiments, temperature-resolved *in situ* SAXS-WAXS experiments, $^{13}$C{$^1$H} ssNMR MAS experiments and 2D $^1$H-$^1$H spin diffusion (NOESY) NMR experiments.
.

**Acknowledgements**

Dr. Andrea Lassenberger and Dr. Sylvain Prévost at Institut Laue Langevin (ILL, Grenoble, France) are kindly acknowledged for their assistance on the SANS experiment. Confocal images were performed at the Institut de Biologie Paris Seine (IBPS) imaging facility. Jean-François Gilles (Imaging Core Facility, Institut de Biologie Paris Seine (IBPS), CNRS, Sorbonne Université, Paris, France) is kindly acknowledged for his assistance on the confocal microscopy imaging. Dr. Guillaume Laurent (LCMCP, Sorbonne Université, Paris, France) is kindly acknowledged for helpful discussions. The French region Ile-de-France SESAME program is acknowledged for financial support (700 MHz NMR spectrometer). Dr. Olivier Diat (ICSM, Marcoule, France), Prof. Aleš Iglič and Prof. Veronika Kralj-Iglič (University of Ljubljana, Slovenia) are kindly acknowledged for helpful discussions.

**Financial support**

WAXS experiments were supported by Soleil Light Source, Saint Aubin, France, proposal N° 20201747. SANS experiments were supported by ILL, proposal N°9-13-778, DOI: 10.5291/ILL-DATA.9-13-778. In situ SAXS-WAXS experiments were supported by ESRF, proposal N° SC-5125.

**TOC Graphic**

Topologically connected nanotube-vesicle above $T_m$ in single bolaform lipid membranes

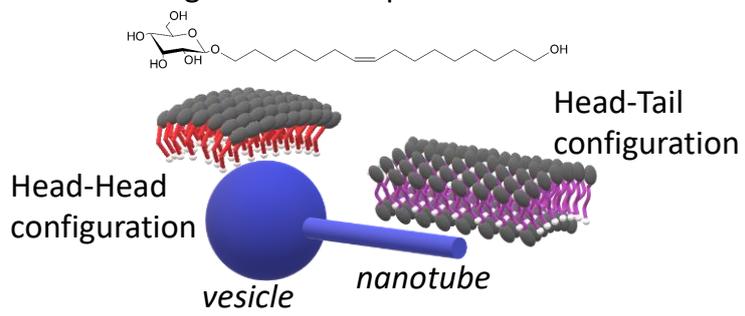

# Supporting Information

# Topological connection between vesicles and nanotubes in single-molecule lipid membranes driven by head-tail interactions

Niki Baccile,[a,*] Cédric Lorthioir,[a] Abdoul Aziz Ba,[a] Patrick Le Griel,[a] Javier Perez,[b] Daniel Hermida-Merino,[c,d] Wim Soetaert,[e] Sophie L. K. W. Roelants[e]

[a] Sorbonne Université, Centre National de la Recherche Scientifique, Laboratoire de Chimie de la Matière Condensée de Paris, LCMCP, Paris, 75005, France

[b] Synchrotron Soleil, L'Orme des Merisiers, Saint-Aubin, BP48, Gif-sur-Yvette Cedex, 91192, France

[c] Netherlands Organisation for Scientific Research (NWO), DUBBLE@ESRF BP CS40220, Grenoble, 38043, France

[d] Departamento de Física Aplicada, CINBIO, Universidade de Vigo, Campus Lagoas-Marcosende, Vigo, 36310, Spain

[e] InBio, Department of Biotechnology, Ghent University, Ghent, 9000, Belgium

**\* Corresponding author:**
Dr. Niki Baccile
E-mail address: niki.baccile@sorbonne-universite.fr
Phone: +33 1 44 27 56 77

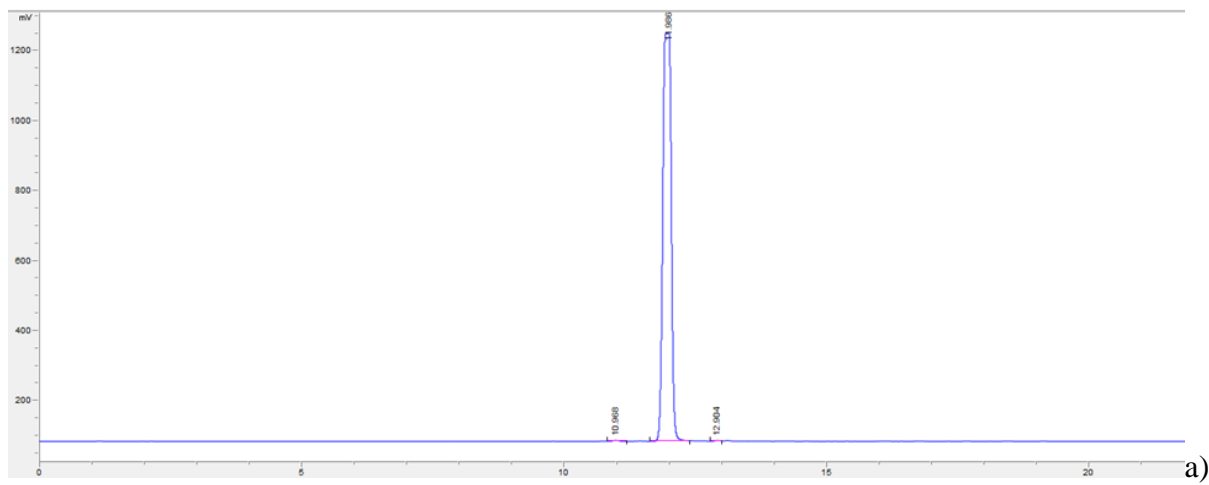

| Parameter | | Method |
|---|---|---|
| Dry Matter (DM %) | 99.6 | Infrared balance 105°C |
| Glucose (%) | 0.12 | HPLC-Metacarb |
| Glycerol (%) | n.d.* | HPLC-Metacarb |
| Free Fatty Acid content (%) | <0.01 | Internal method BBEPP: GC with FID Detector |
| Oil (%) | <0.01 | Internal method BBEPP: GC with FID Detector |
| Moulds (CFU/g) | <10 | 3M-Nordval n°16 B |
| Yeasts (CFU/g) | <10 | 3M-Nordval n°16 B |
| Anaerobic count (CFU/g) | <10 | SP-VG M005 B |
| Protein (%) | t.b.d.* | BCA protein assay |

**Figure S 1 - a) HPLC-ELSD chromatogram and composition table.**

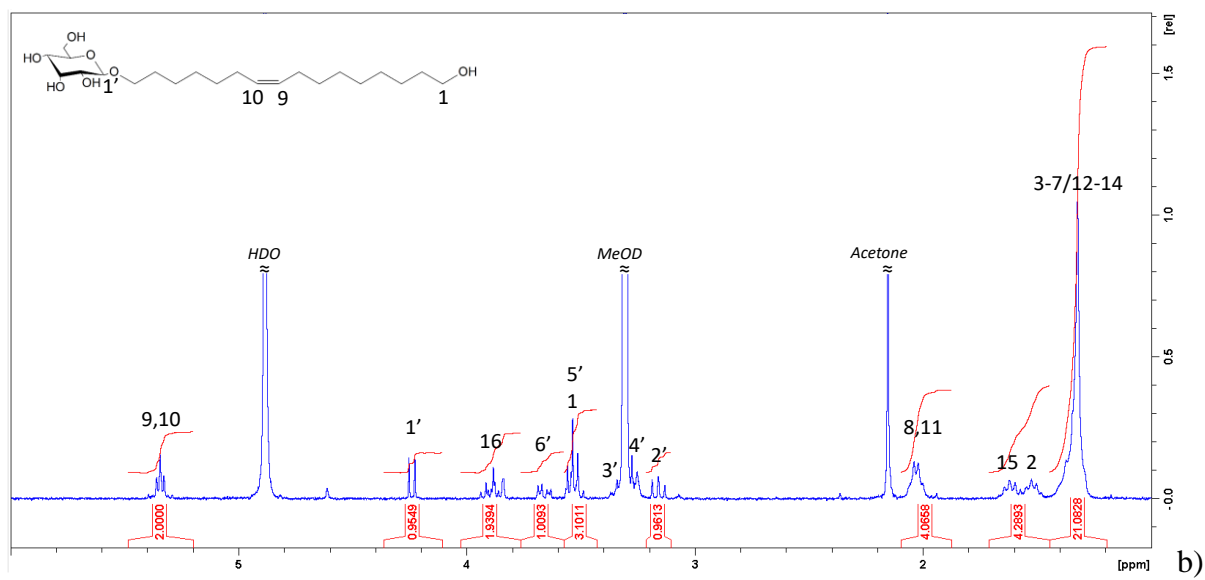

| Group | N° | δ / ppm |
|---|---|---|
| CH₂ | 1 | 3.54 |
| CH₂ | 2 | 1.53 |
| CH₂ | 3-7/12-14 | 1.32 |
| CH₂ | 8,11 | 2.03 |
| CH | 9,10 | 5.34 |
| CH₂ | 15 | 1.62 |
| CH₂ | 16 | 3.88 |
| CH | 1' | 4.24 |
| CH | 2' | 3.16 |
| CH | 3' | 3.34 |
| CH | 4' | 3.29 |
| CH | 5' | 3.54 |
| CH₂ | 6' | 3.66 |

**Figure S 1 (next) – b)** $^1$H NMR spectrum (MeOD-d4) recorded on the purified G-C18:1-OH compounds and related assignment

S3

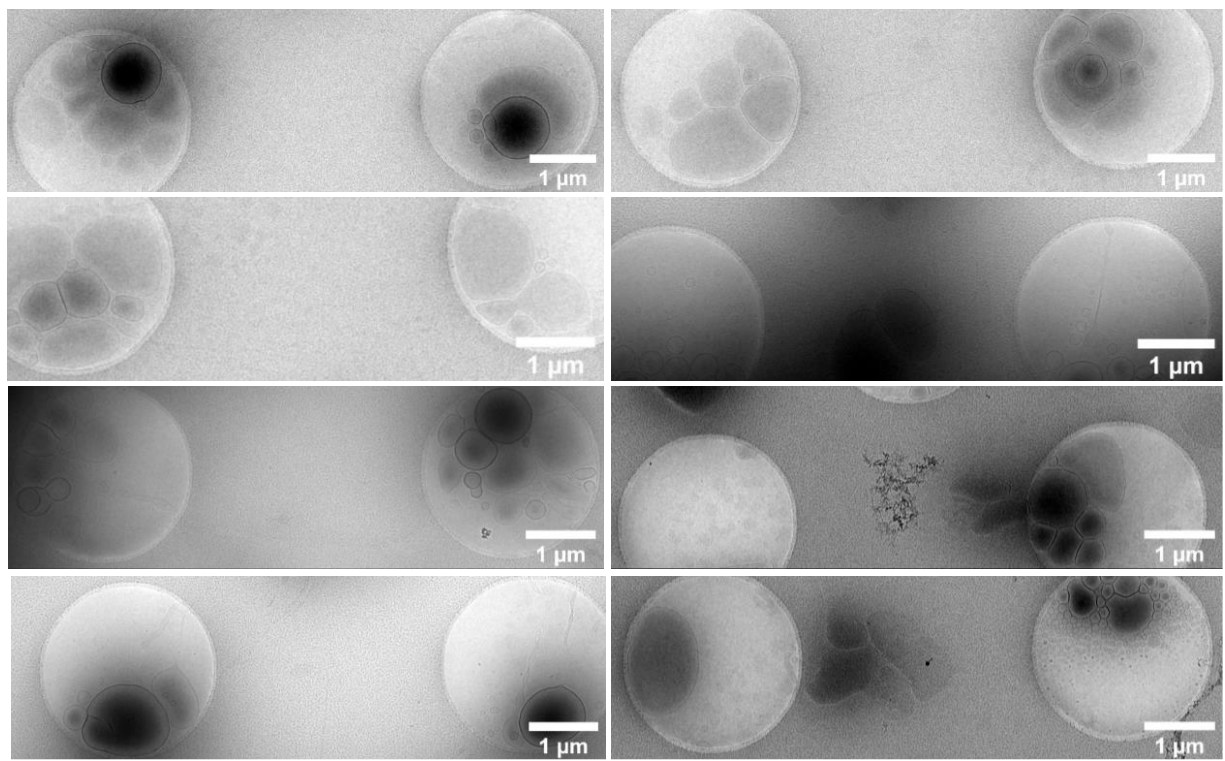

**Figure S 2 – Cryo-TEM images recorded for a 5 mg/mL G-C18:1-OH aqueous solution heat at *T*\*= 130°C.**

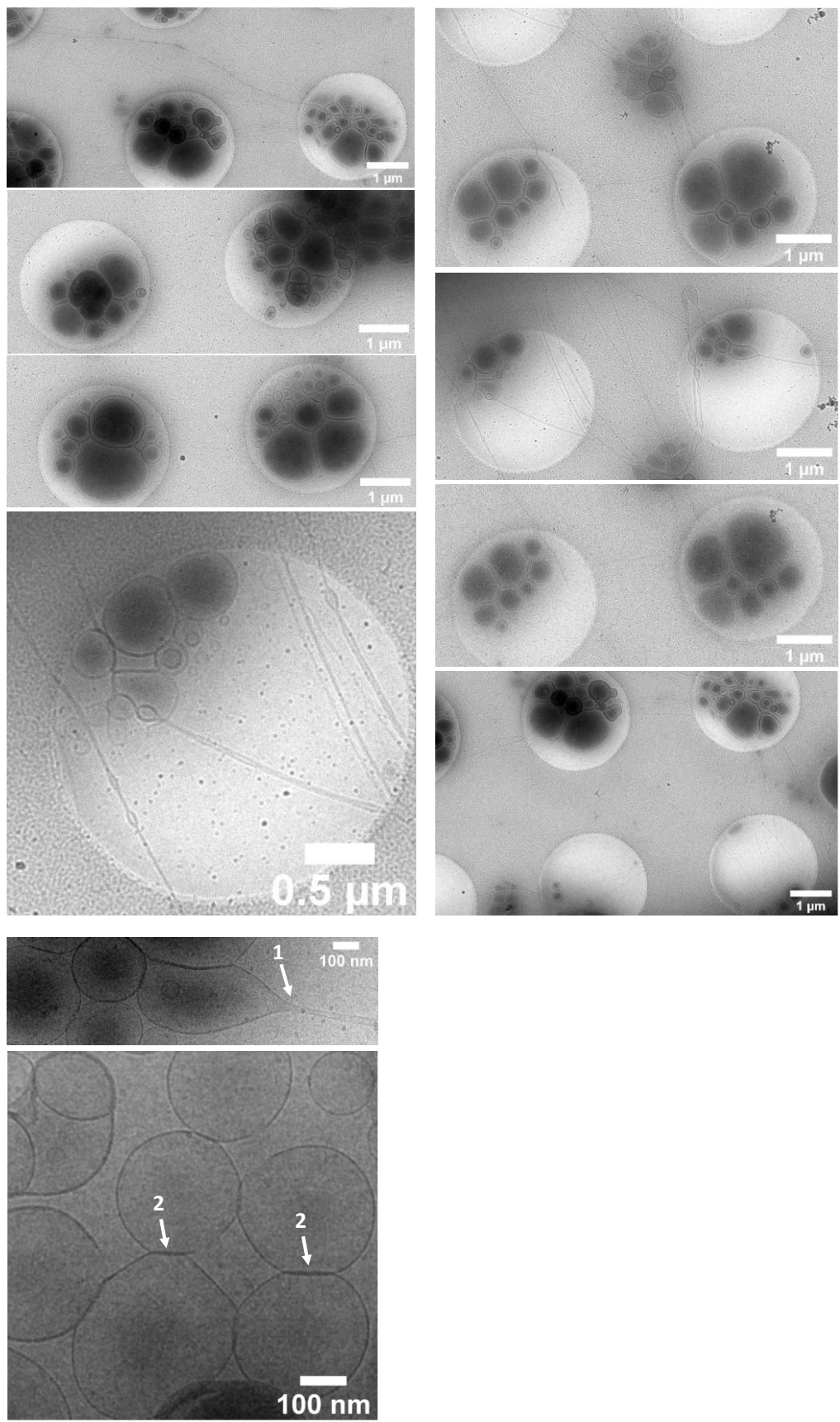

**Figure S 3 - Cryo-TEM images recorded for a 5 mg/mL G-C18:1-OH aqueous solution heat at $T^* = 90°C$.**

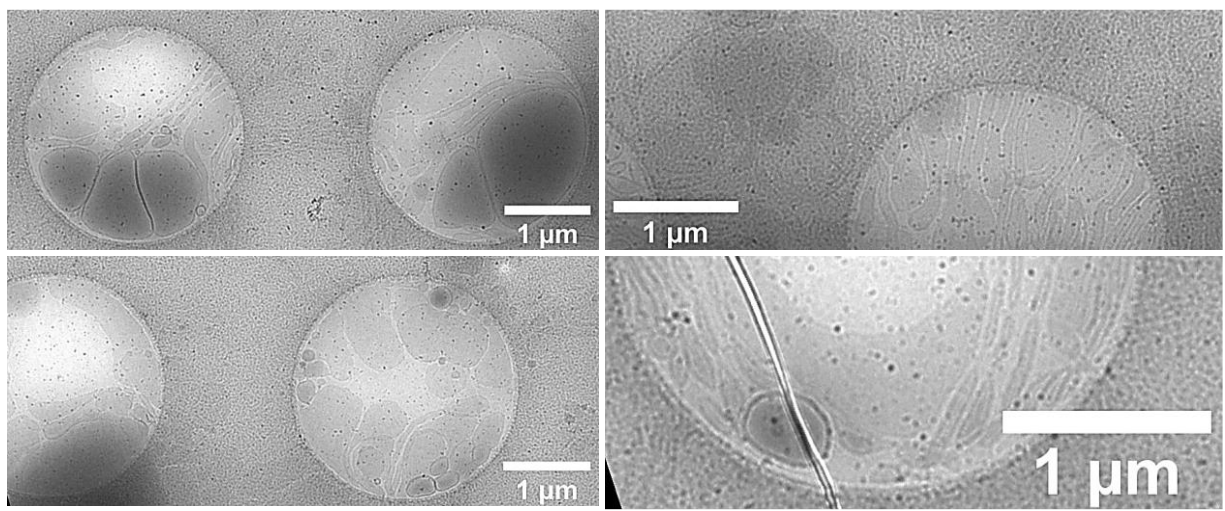
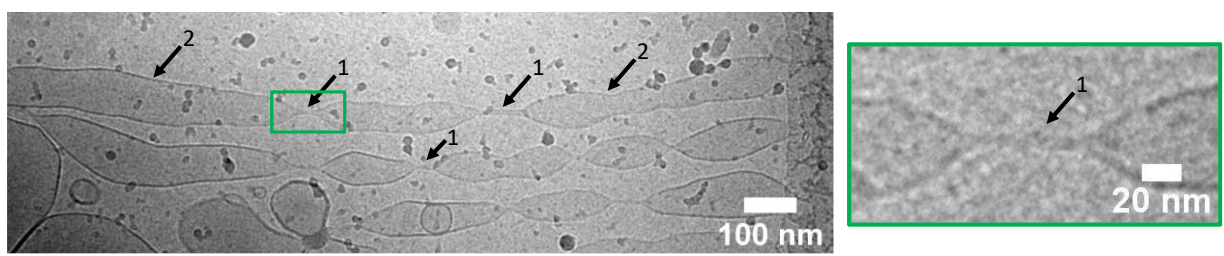
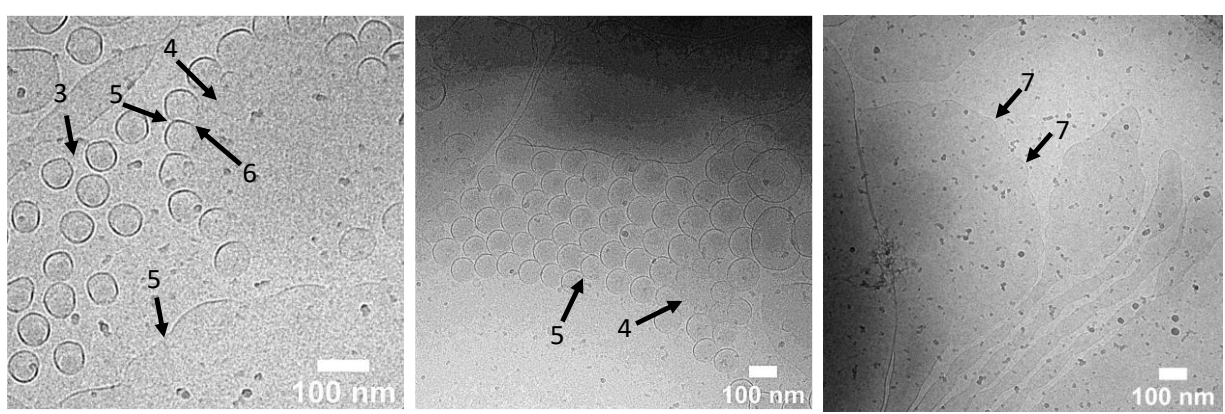

**Figure S 4 – Cryo-TEM images recorded for a 5 mg/mL G-C18:1-OH aqueous solution heat at $T^*= 70°C$.**

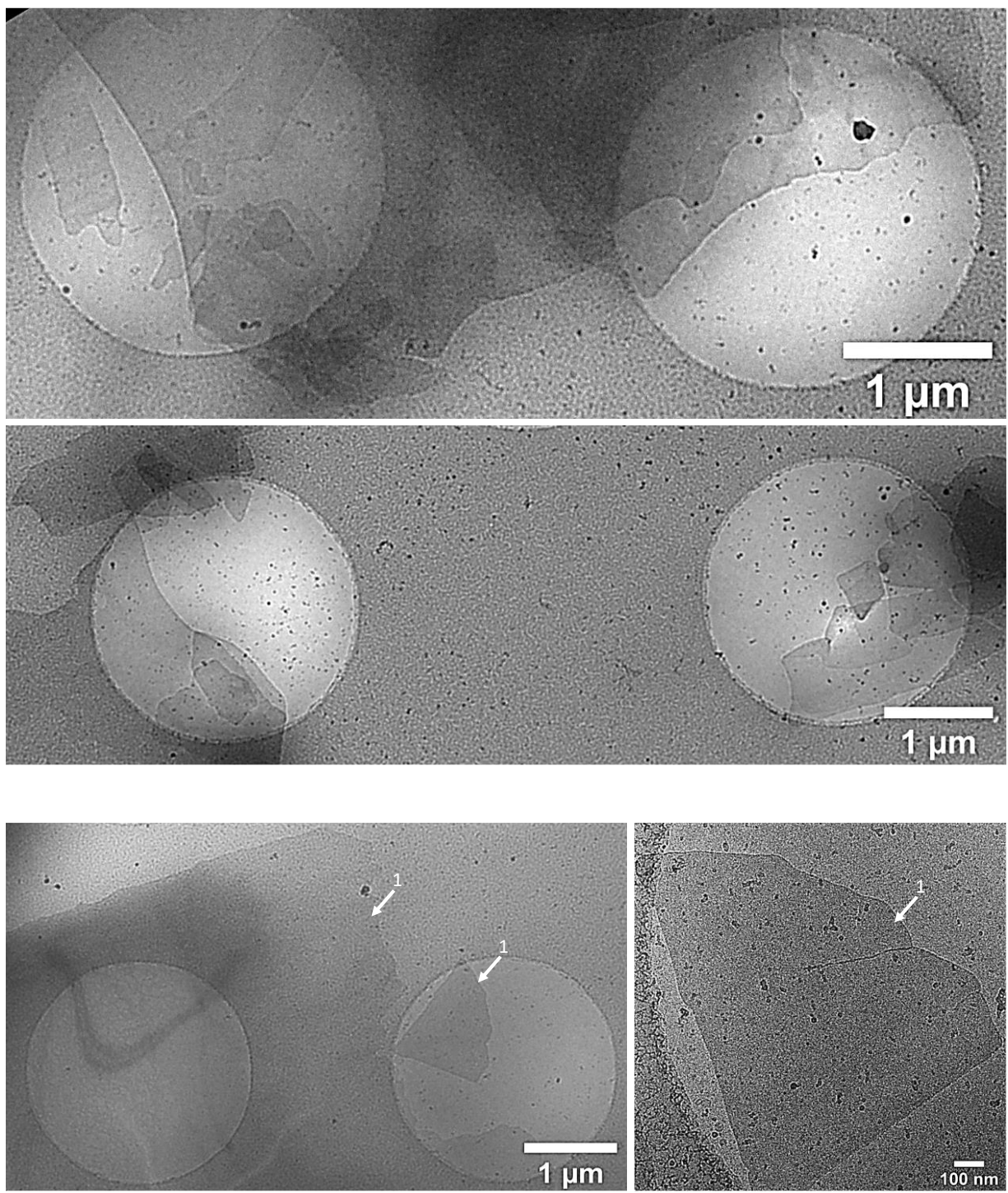

**Figure S 5 – Cryo-TEM images recorded for a 5 mg/mL G-C18:1-OH aqueous solution heat at *T\**= 25°C.**

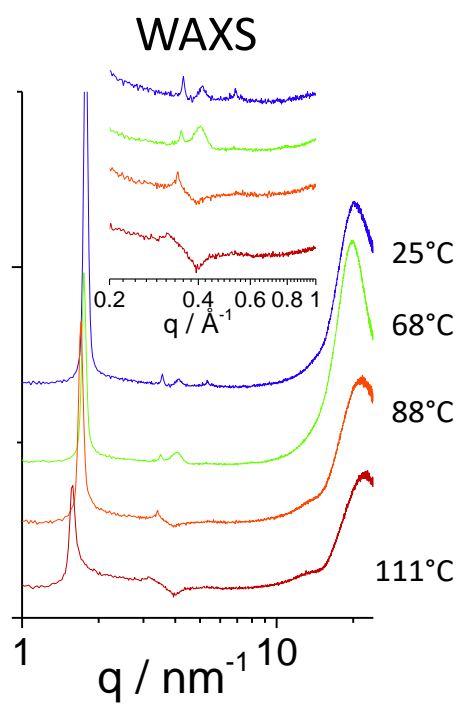

**Figure S 6 –WAXS experiments recorded on aqueous solutions of G-C18:1-OH at 5 mg/mL.**

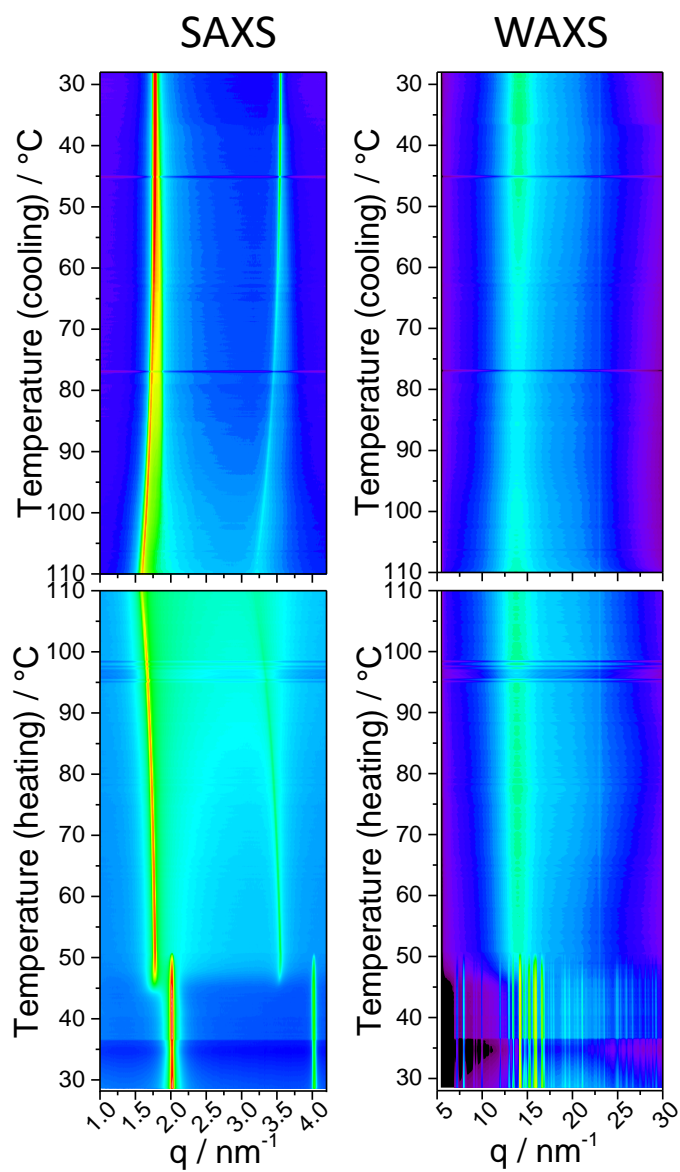

**Figure S 7 – Temperature-resolved *in situ* SAXS-WAXS experiments recorded on a G-C18:1-OH sample at 50 wt% in water. The sample is sealed in a borosilicate 2 mm capillary and undergoing a heating-cooling cycle (rate: 1°C/min) between room temperature and 110°C using a Linkam stage set in front of the X-ray beam.**

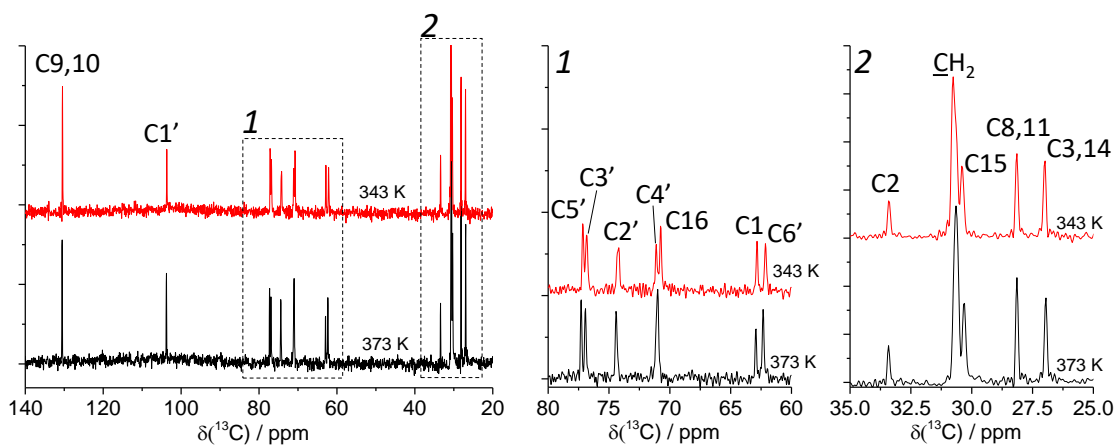

| Position | δ / ppm | | |
|---|---|---|---|
| | $^{13}$C Liquid (DMSO-d6) | $^{13}$C solid state (T= 110°C) | $^{1}$H solid state (T= 110°C) |
| **9,10** | 130.09 | 130.46 | 5.56 |
| **1'** | 103.29 | 103.69 | 4.602 |
| **5'** | 77.24 | 77.21 | 3.734 |
| **3'** | 77.24 | 76.85 | 3.665 |
| **2'** | 73.88 | 74.32 | 3.553 |
| **4'** | 70.52 | 71.03 | 4.086 |
| **16** | 69.00 | 70.89 | 3.732 |
| **1** | 61.53/61.17 | 62.86 | 3.809 |
| **6'** | 61.53/61.17 | 62.25 | 4.108 |
| **2** | 33.00 | 33.34 | 1.879 |
| **15** | 29.41 | 30.24 | 1.971 |
| **-(CH$_2$)-** | 29.41 | 30.58 | 1.645 |
| **8,11** | 27.06 | 28.08 | 2.334 |
| **3,14** | 25.97 | 26.89 | 1.665 |

**Figure S 8 – Single pulse high-power decoupling $^{13}$C{$^{1}$H} ssNMR MAS (2 kHz) experiments recorded on a 50 wt% G-C18:1-OH sample prepared in D$_2$O. Spectra are recorded at 110°C and at 70°C after cooling from 110°C. Attribution of each peak position for both $^{13}$C and $^{1}$H in the solid state at 110°c is given in the table and compared to the chemical shift position in the liquid.**

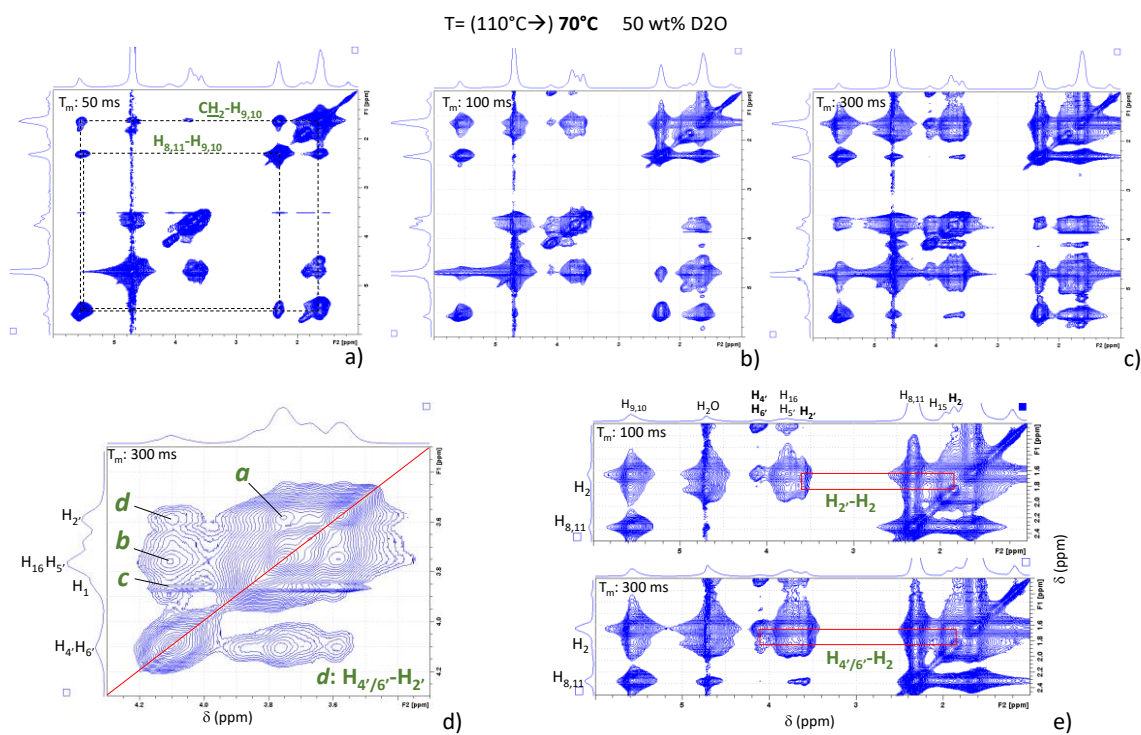

**Figure S 9** – 2D $^1$H-$^1$H spin diffusion (NOESY) NMR experiments using a) 50 ms, b) 100 ms and c) 300 ms of mixing time recorded on a 50 wt% G-C18:1-OH sample in D$_2$O. Sample temperature is regulated at 70°C after cooling from 110°C. Highlights of the glucose and glucose-aliphatic regions are shown in d) and e), respectively (mixing time is reported on each contour plot).

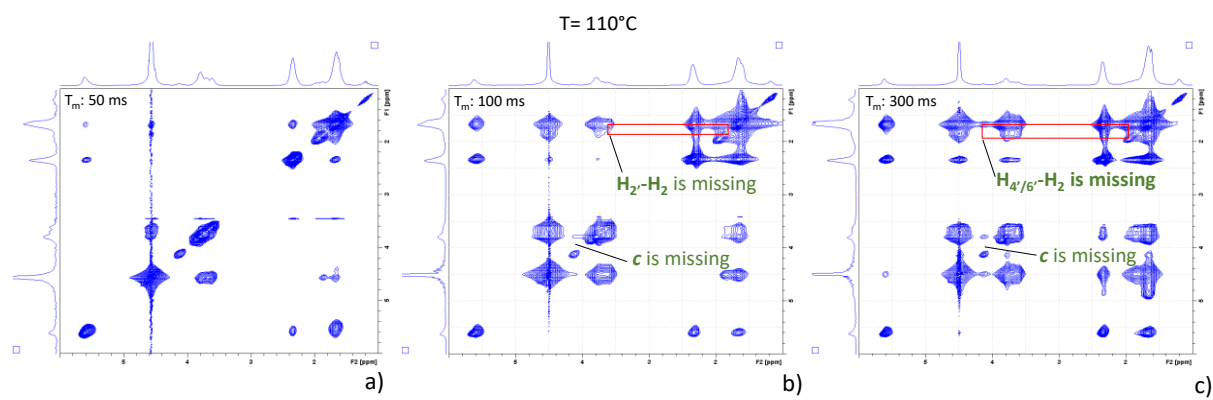

**Figure S 10** – 2D $^1$H-$^1$H spin diffusion (NOESY) NMR experiments using a) 50 ms, b) 100 ms and c) 300 ms of mixing time recorded on a 50 wt% G-C18:1-OH sample in D$_2$O. Sample temperature is regulated at 110°C.

**Comments on 2D spin diffusion experiments performed at 70°C (Figure S 9) and 110°C (Figure S 10) at 50 ms.**

Experiments performed at mixing time of 50 ms (Figure S 9a, Figure S 10a) show only the correlation between close neighbors: the cross-peaks at 130.46 and 30.58 ppm, or at 130.46 and 28.08 ppm, identify H$_{9,10}$-(C$\underline{H}_2$)$_n$ and H$_{9,10}$-H$_{8,11}$ correlations, respectively. All other identifiable cross-peaks also show short-range interactions (e.g., CH-CH in glucose). At 300 ms, the number of cross-peaks becomes higher, both at 70°C and 110°C, and they also identify long range interactions, like $H_{2'}$-$H_2$ and $H_{4'/6'}$-$H_2$ in Figure S 9e. In this regard, the mixing time of 100 ms represents a good compromise between short and long-range interactions. At 70°C, one can observe the $H_{2'}$-$H_2$ cross-peak but not the $H_{4'/6'}$-$H_2$ cross-peak, while at 110°C, one cannot observe neither one nor the other. This indicates that the $H_{2'}$-$H_2$ correlation is medium-range at 70°C and long-range at 110°C. Similar considerations occur for the glucose region, and in particular for the cross-peaks *a*, *b*, *c*, *d* observed at 300 ms at 70°C (Figure S 9d), while only *b* is detected at 110°C at the same mixing time.